%% file: main.tex
\documentclass[letterpaper,11pt]{article}
\usepackage{jcappub} 

\usepackage{lineno}
\usepackage{xspace}
\usepackage{xcolor}
\usepackage{amsmath}
\usepackage{mathtools}
\usepackage[margin=2.5cm]{geometry}
\usepackage{orcidlink}
\usepackage{aas_macros}

\defcitealias{Riess_2022_SH0ES}{R22a}
\defcitealias{SH0ES_ClusterCeph}{R22b}
\defcitealias{Planck_2018}{Planck 2018}
\defcitealias{Blondin_2012_CFAspec}{Blondin et al. (2012)}
\defcitealias{Matheson_2008_CFAspec_indiv}{Matheson et al. (2008)}
\defcitealias{Stahl_2020_BSNIP}{Stahl et al. (2020)}
\defcitealias{Silverman_2012_BSNIP1}{Silverman et al. (2012)}
\defcitealias{Folatelli_2013_CSPspec}{Folatelli et al. (2013)}
\defcitealias{Sako_2018_SDSS-SN-DR}{Sako et al. (2018)}
\defcitealias{Dettman_2021_FoundationSpec}{Dettman et al. (2021)}
\defcitealias{Kenworthy_2021_SALT3}{Kenworthy et al. (2021)}
\defcitealias{Pan_2022_PS1Spec}{Pan et al. (2022)}
\defcitealias{Guillochon_2017_OSC}{Giollochon et al. 2017}
\defcitealias{Boone_2021_TwinsEmbeddinng_I}{B21a}
\defcitealias{Boone_2021_TwinsEmbeddinng_II}{B21b}

\newcommand{\deepSIP}{\texttt{deepSIP}\xspace}
\newcommand{\Hnaught}{$H_0$\xspace}
\newcommand{\dmfifteen}{$\Delta m_{15}$\xspace}

\graphicspath{{figures/}}

\title{Leveraging SN~Ia spectroscopic similarity to improve the measurement of $H_0$}

\author[1]{Yukei S. Murakami\orcidlink{0000-0002-8342-3804},}
\author[1,2]{Adam G. Riess\orcidlink{0000-0002-6124-1196},}
\author[3]{\\Benjamin E. Stahl\orcidlink{0000-0002-3169-3167},}
\author[4]{W. D'Arcy Kenworthy\orcidlink{0000-0002-5153-5983},}
\author[5,6]{\\Dahne-More A. Pluck\orcidlink{0000-0003-3724-4773},}
\author[1]{Antonella Macoretta\orcidlink{0000-0002-5874-4892},}
\author[7]{\\Dillon Brout\orcidlink{0000-0001-5201-8374},}
\author[8]{David O. Jones\orcidlink{0000-0002-6230-0151},}
\author[9]{\\Dan M. Scolnic\orcidlink{0000-0002-4934-5849},}
\author[3]{and Alexei V. Filippenko\orcidlink{0000-0003-3460-0103}}

\affiliation[1]{\small Department of Physics and Astronomy, Johns Hopkins University, Baltimore, MD 21218, USA}
\affiliation[2]{\small Space Telescope Science Institute, 3700 San Martin Drive, Baltimore, MD 21218, USA}
\affiliation[3]{\small Department of Astronomy, University of California, Berkeley, CA 94720-3411, USA}
\affiliation[4]{\small The Oskar Klein Centre for Cosmoparticle Physics, Department of Physics, Stockholm University, SE-10691 Stockholm, Sweden}
\affiliation[5]{\small Rowland Physics Summer Research Fellow}
\affiliation[6]{\small Department of Physics, University of Maryland, Baltimore County, Baltimore, MD 21250, USA}
\affiliation[7]{\small Center for Astrophysics, Harvard \& Smithsonian, 60 Garden St., Cambridge, MA 02138, USA}
\affiliation[8]{\small Gemini Observatory, NSF's NOIRLab, 670 N. A'ohoku Place, Hilo, Hawai'i, 96720, USA}
\affiliation[9]{\small Department of Physics, Duke University, Durham, NC 27708, USA}

\emailAdd{ymuraka2@jhu.edu}

\abstract{
Recent studies suggest spectroscopic differences explain a fraction of the variation in Type Ia supernova (SN~Ia) luminosities after light-curve/color standardization. 
In this work, (i) we empirically characterize the variations of standardized SN~Ia luminosities, and (ii) we use a spectroscopically inferred parameter, SIP, to improve the precision of SNe~Ia along the distance ladder and the determination of the Hubble constant ($H_0$). First, we show that the \texttt{Pantheon+} covariance model modestly overestimates the uncertainty of standardized magnitudes by $\sim 7$\%, in the parameter space used by the \texttt{SH0ES} Team to measure $H_0$; accounting for this alone yields $H_0 = 73.01 \pm 0.92$\,km\,s$^{-1}$\,Mpc$^{-1}$. Furthermore, accounting for spectroscopic similarity between SNe~Ia on the distance ladder reduces their relative scatter to $\sim0.12$\,mag per object (compared to $\sim 0.14$\,mag previously). Combining these two findings in the model of SN covariance, we find an overall 14\% reduction (to $\pm 0.85$\,km\,s$^{-1}$\,Mpc$^{-1}$) of the uncertainty in the Hubble constant and a modest increase in its value.
Including a budget for systematic uncertainties itemized by Riess et al. (2022a), we report an updated local Hubble constant with $\sim1.2$\% uncertainty, $H_0 = 73.29 \pm 0.90$\,km\,s$^{-1}$\,Mpc$^{-1}$. We conclude that spectroscopic differences among photometrically standardized SNe~Ia do not explain the ``Hubble tension." Rather, accounting for such differences increases its significance, as the discrepancy against $\Lambda$CDM calibrated by the {\it Planck} 2018 measurement rises to 5.7$\sigma$.
}

\begin{document}
\maketitle
\flushbottom

\section{Introduction} \label{sec:intro}
    \subsection{Current affairs of SN~Ia cosmology}
    Type Ia supernovae (SNe~Ia) --- violent explosions of white dwarf stars in multistar systems --- have been used as cosmological distance indicators for nearly three decades, and were critical to  the discoveries that led to the current cosmological model \citep[][]{Riess_1998,Perlmutter_1999}. Their utility as distance indicators is enabled by their large and relatively homogeneous peak luminosities that can be further standardized via photometric time-series observations \citep[for a review, see][]{Branch_Wheeler_2017_SNtextbook}.
    The process of calibrating a specific SN~Ia's luminosity (from the small dispersion that such objects have) is called standardization, and is typically accomplished by exploiting empirical relationships between photometric properties such as light-curve morphology and color. There is a mature literature on SN~Ia standardization techniques, with notable methods including template fitter \citep{Hamuy_1996_SNtemplatefit}, \texttt{MLCS} \citep{Riess_1996_MLCS}, \texttt{MLCS2k2} \citep{Jha_2007_MLCS2k2}, \texttt{SALT2} \citep[][]{Guy_2007_SALT2,Guy_2010_SNLS}, \texttt{SALT3} \citep{Kenworthy_2021_SALT3}, \texttt{SiFTO} \citep{Conley_2008_SiFTO}, \texttt{SNooPy} \citep{Burns_2011_SNooPy}, \texttt{SNEMO} \citep{Saunders_2018_SNEMO}, \texttt{SUGAR} \citep{Leget_2020_SUGAR}, and \texttt{BayesSN} \citep{Mandel_2011_BayesSN1,Mandel_2022_BayesSN2}.
    Most of these methods follow the ``Phillips relation" \citep[][]{Phillips_1993_PhillipsRelation} --- i.e., that the maximum luminosity of an SN~Ia correlates with the rate of decline of its luminosity (as measured from an optical light curve).  Many also include an astrophysical extinction model in which redder SNe~Ia appear fainter \citep{Riess_1996_MLCS}.  It is worth noting that past cosmological inferences from SNe~Ia have not been particularly sensitive to which of these methods are employed, and most are capable of reducing the observed dispersion of SNe~Ia from an apparent range of 0.4--0.5\,mag to a range of 0.1--0.2\,mag, thereby increasing their cosmological leverage by an order of magnitude.

    A comparably small amount of post-standardization luminosity variation, 0.05--0.1\,mag, may be correlated with SN host-galaxy properties (with some dependence on the standardization method and sample). In particular, the host's stellar mass \citep[e.g.,][]{Kelly_2010_SNmass,Gupta_2011_SNmass,Childress_2013_SNmass,Uddin_2017_SNmass,Jones_2018_SNmass}, star-formation rate \citep[SFR; e.g.,][]{Rigault_2020_SNsfr,Briday_2022_SNsfr}, population age \citep[e.g.,][]{Kang_2020_SNage,Zhang_2021_SNage}, morphology \citep[e.g.,][]{Pruzhinskaya_2020_SNmorph}, color \citep{Kelsey_2021_SNcolor}, and cosmic mean metallicity \citep[][]{Li_2021_SNmetal} have all been explored.
    \cite{Murakami_2021_SNeLuminosity} showed that such parameterizations are all consistent with variation of galaxy properties and that the precision of luminosity corrections depends on the accuracy of the measurements or estimation of the chosen parameter. Most recently, \cite{Brout&Scolnic_2021} demonstrated that apparent SN~Ia differences that would otherwise appear to correlate with the previously listed host properties may instead be explained by differences in the reddening laws of hosts of different masses. Their analysis finds that modeling SN~Ia scatter driven by variation in the dust attenuation law ($R_V$) removes the variation on other parameters (e.g., the ``mass step'') and minimizes the standardized luminosity after bias correction (following the  \texttt{BBC} framework described by \cite[][]{Kessler&Scolnic_2017_BBC}).  
    
    A further innovation has been the compilation of large numbers of SNe~Ia that are ``cross-calibrated'' through the use of large sky surveys and common reference stars \citep{Scolnic_2021_P+data,Betoule_2014_JLA,Brout_2021_fragilistic,Rubin_2022_crosscal}.  Lower dispersion can be realized for a single, well-calibrated survey due to the photometric (or spectroscopic) consistency of all measurements within the sample, but a single survey is otherwise impractical for the measurement of cosmological parameters owing to the need to span different distance ranges.  Because SNe~Ia within range of primary distance indicators (i.e., resolved stellar populations) are rare and thus require decades to accumulate, the effort to measure $H_0$ must necessarily leverage such compilations rather than any single survey.

    By combining these empirical techniques --- i.e., photometric standardization, scatter modeling, and sample calibration--- state-of-the-art cosmological studies based on large-sample, multidecade SN~Ia compilations achieve a scatter as low as $\sim0.14$\,mag at low redshift ($z<0.15$; Riess et al. 2022a \cite[][hereafter \citetalias{Riess_2022_SH0ES}]{Riess_2022_SH0ES}) and $\sim 0.17$\,mag at high redshift ($z<2.26$; \cite{Brout_2022_P+Cosmo}).  
    The uncertainty per SN~Ia post-standardization is estimated from scatter modeling at fixed redshift and it sets the scale for the error in the mean of the sample of 42 SNe~Ia used to calibrate their fiducial luminosity.  This error term (i.e., the uncertainty in the mean of the locally-calibrated sample) accounts for 60--70\% of the \Hnaught uncertainty.
    It is therefore of great importance to (i) understand its origin and (ii) seek to reduce the size of unexplained post-standardization luminosity scatter, if possible.

    \subsection{The use of spectra for standardization}
    Since the rise of SNe~Ia as cosmological distance indicators, the success of photometric standardization methods has motivated studies to use spectroscopic information in attempts to model yet-to-be-explained intrinsic variations that photometric standardization cannot correct, as well as to tie the intrinsic variations with the objects' underlying explosion mechanism(s). 
    \cite{Nugent_1995} first showed a significant correlation between spectral features (ratio of fractional depths of a pair of Ca~II and Si~II absorption lines) and the absolute magnitudes of SNe~Ia.
    An in-depth analysis of absorption lines was presented by \cite{Branch_2006}, in which the sample SNe were grouped into four possible spectral subtypes. The extent of the differences in the variations of absolute magnitude ($M_B$) between each subtype have been presented in an extensive study of spectroscopic variations by \cite{Blondin_2012_CFAspec}, and several important relations between SN~Ia spectra, luminosity decline rate ($\Delta m_{15}(B)$), and $M_B$ have been confirmed: (i) different subtypes, though the distribution is rather continuous and classification boundaries are not clear, have different $\Delta m_{15}(B)$--$M_B$ slopes, (ii) a single parameter is not enough to capture the diversity, and (iii) the relation between spectroscopic parameters and $M_B$ may be nonlinear.  
    
    Although efforts are still underway to tie the exact physical mechanism to specific observable features in SN~Ia spectra \citep[e.g.,][]{Siebert_2019_kaepora,Siebert_2020_ejectaVelocity,Zhang_2020_ejectavelocity,Pan_2022_PS1Spec}, recent literature has shown that an empirical and arbitrary parameterization of spectra at the time of maximum brightness can further reduce the scatter in SN~Ia luminosity. \cite[][]{Fakhouri_2015_Twins} introduced the ``twins'' approach, which performs pairwise comparisons of SN~Ia spectra (taken near maximum or with interpolated time series) and ranks their similarities. As previous studies suggested, it was shown that \emph{spectroscopically similar SNe~Ia have similar luminosities}, yielding an intrinsic scatter as low as 0.06--0.07\,mag (although this is based on a relatively small sample from a single photometric survey, which already reduces dispersion over compilations by eliminating residual calibration errors).
    Boone et al. 2021a \cite[][hereafter \citetalias{Boone_2021_TwinsEmbeddinng_I}]{Boone_2021_TwinsEmbeddinng_I} updated this ``twins'' analysis by taking advantage of a nonlinear parameterization (rather than relative pairwise comparison) based on a manifold learning technique which yields an arbitrary number of parameters to empirically characterize variations of spectra. The authors show that three parameters (presented as $\eta_1$, $\eta_2$, and $\eta_3$) capture the intrinsic variation at the time of maximum brightness\footnote{\citetalias{Boone_2021_TwinsEmbeddinng_I} prepares the dereddened (i.e., representative of intrinsic) spectra at the time of maximum using the ``Reading Between the Lines'' (RBTL) technique, which simultaneously finds the wavelength-dependent intrinsic diversity to only use the wavelengths with small diversity for dereddening.} most efficiently and claims to explain a large fraction of the intrinsic diversity. An implementation of this technique in a cosmological context is demonstrated by \cite{Boone_2021_TwinsEmbedding_II}, where the authors utilize Gaussian processes (GP) to map the variations captured by three parameters to standardized luminosity. This procedure produces a dispersion of standardized absolute magnitudes \emph{for a single, specific survey} of $\sim 0.084$\,mag, which could be reduced to $\sim 0.073$\,mag when a larger sample is used.

    In the context of a full and realistic cosmological analysis, however, there are challenges involved with directly employing the ``twins'' technique.
    First, it is necessary to construct a large and highly uniform sample of spectral and photometric data to perform cosmological analyses. Combining (by necessity) the multiple surveys, telescopes, and observation methods necessitates the cross-calibration of instruments \citep[e.g.,][]{Brout_2021_fragilistic} and simulation-based corrections that account for selection bias \citep[e.g.,][]{Kessler&Scolnic_2017_BBC,Brout&Scolnic_2021}. 
    More importantly, spectra of such SNe often come from a variety of instruments even within each photometric survey, and this diversity necessitates an even more technically involved, wavelength-dependent, possibly nonlinear cross-calibration of fluxes across all instruments used, including those that may no longer exist.
    This is an issue peculiar to the determination of \Hnaught: while it is possible to design a new survey to refine the expansion history of the universe at $z>0.02$, the low rate of SN~Ia discoveries within the volume where reliable stellar distances are available (i.e., $z<0.01$) necessitates the use of existing, decades-old photometry.  For these SNe~Ia, a uniform set of spectrophotometric time-series spectra over the same range of wavelength and spectral features simply does not exist.

    To address these challenges while still pursuing gains from spectroscopic similarity in the full cosmological analysis of \Hnaught, we leverage an existing framework (i.e., SALT2+BS21) with an analysis of normalized, rectified spectra to determine whether the use of more limited spectroscopic information can still reduce variation.
    In particular, we combine the most recent baseline SH0ES analysis from Riess et al. (2022b) \cite{SH0ES_ClusterCeph} (hereafter \citetalias{SH0ES_ClusterCeph}) with a machine-learning-based spectroscopic analysis framework \citep[\deepSIP;][]{Stahl+2020_deepSIP} and analyze the nonlinear relation between the proximity in parameter space for a pair of SNe and their luminosity covariance. Using \deepSIP (which, at its core, is a convolutional neural network that parameterizes the information embedded in the spectrum of an SN~Ia) allows us to, for the first time\footnote{We note that \cite{Riess_2022_SH0ES} used \deepSIP\citep[][]{Stahl+2020_deepSIP} to demonstrate that all 42 calibrator SNe~Ia are typical of the SN~Ia spectroscopic class but did not employ this measure quantitatively for improving the measurements.}, introduce spectroscopic similarity information into the distance ladder leveraged for the $H_0$ measurement. We find that incorporating this information into the SH0ES analysis increases the precision of SN~Ia distance information by $\sim 14\%$, equivalent to an increase in the number of calibrators by $\sim 30\%$ \emph{without} requiring any further data collection.

    A newly assembled spectroscopic dataset belonging to SH0ES SNe~Ia is described in Sec.~\ref{sec:data} and the \deepSIP-inferred properties of its constituents is explored in Sec.~\ref{sec:deepSIP}.
    Using the obtained spectroscopic parameter, we measure the size of the SN~Ia luminosity scatter explained by the principle of spectral similarity in Sec.~\ref{sec:SP} and \ref{sec:SP_implication}. We then implement a kernel-based model to include such information in the cosmological distance ladder in Sec.~\ref{sec:COV}. In Sec.~\ref{sec:distladder}, we show that our new model improves the uncertainty on \Hnaught by $\sim 14$\%. 
    We discuss the effectiveness of the spectroscopic parameterization using \deepSIP and possible improvements in Sec.~\ref{sec:discussions}.
    A summary of our findings and a discussion of further possible improvements are provided in Sec.~\ref{sec:conclusion}.

\section{Data} \label{sec:data}
    \input{spec_sources}
    \subsection{Cosmological SN~Ia datasets: \texttt{Pantheon+} and \texttt{SH0ES}}
    \label{sec:data-cosmologydata}
    We use the photometric parameters provided in the \texttt{Pantheon+} analysis \citep{Scolnic_2021_P+data}. \texttt{Pantheon+} is the largest compilation of SNe~Ia which are viable for a cosmological study, and which span four decades of observations across numerous surveys and individual campaigns. The photometric data are cross-calibrated using wide-angle surveys \citep{Brout_2021_fragilistic} and parametrized by \texttt{SALT2} \citep{Guy_2007_SALT2}. In our analysis, the relevant parameters are the apparent maximum brightness ($m_B$), stretch ($x_1$), and color ($c$). As briefly discussed in Sec.~\ref{sec:intro}, the combination of different observation methods and instruments leads to nontrivial biases and covariance, the corrections for and proper treatment of which can have a significant impact on the results of cosmological analyses \citep[][]{Kessler&Scolnic_2017_BBC}. The \texttt{Pantheon+} dataset is processed using the BEAMS with bias correction (BBC) method and the \texttt{BS21} model (\cite{Brout&Scolnic_2021}; improved by \cite{Popovic_2021_BS21update}, hence the \texttt{BS21/P21} model hereafter). We note that the color-dependent scatter introduced by the \texttt{BS21/P21} model has a consequence in our modeling, and will therefore be discussed in Sec.~\ref{sec:COV}.

    The \texttt{SH0ES} dataset (\citetalias{Riess_2022_SH0ES}) is comprised of a subset of \texttt{Pantheon+} SN~Ia measurements along with Cepheid-derived distances to some of their host galaxies, thus allowing for their absolute distances to be calibrated. The dataset contains 42 Cepheid-calibrated nearby SNe~Ia (SNe-CC) and 277 Hubble-flow SNe~Ia (SNe-HF), from which the Hubble constant can be constrained.
    In the most recent update from the SH0ES team, \citetalias{SH0ES_ClusterCeph} have used cluster Cepheid parallax observations from Gaia DR3 to reduce the uncertainty in their $H_0$ measurement to below 1\,km\,s$^{-1}$\,Mpc$^{-1}$. The distance ladder evaluated in Sec.~\ref{sec:distladder} shares identical observational data with \citetalias{SH0ES_ClusterCeph}, with an update to its covariance matrix which we prepare in Sec.~\ref{sec:COV}.

    \subsection{Spectra}
    \label{sec:data-spectra}
    We collect spectra of the SNe~Ia in the \texttt{SH0ES} baseline sample primarily from eight published data sources (see Table~\ref{tab:spec_sources}). The assembly of spectra includes the data used by \citetalias{Riess_2022_SH0ES} to confirm that the 42 Cepheid Calibrator SNe are ``normal'' types.
    The source publications are mostly the official data releases of the main surveys that constitute the \texttt{SH0ES} baseline sample, including
    CFA \citep{Riess_1999_CFA1,Jha_2006_CFA2,Hicken_2009_CFA3,Hicken_2012_CFA4}, LOSS/BSNIP \citep{Filippenko_2001_KAIT,Mo_2010_LOSS1_photIa,Stahl_2019_LOSS2_photIa}, CSP \citep{Hamuy_2006_CSP}, SDSS \citep{Sako_2008_SDSS-SN}, Foundation \citep{Foley_2018_Foundation}, and PS1 \citep{Scolnic_2018_PS1phot}.
    Since we use normalized spectra and our measurements are not dependent on survey characteristics, we do not separate photometric duplicates (i.e., individual SNe~Ia observed by multiple surveys, consistent with \texttt{Pantheon+}'s treatment). When spectra for a single SN are available from multiple surveys, we retain all spectra except for instances where one spectrum is numerically identical to another.

\section{Estimating \textit{Spectroscopically Inferred Parameter (SIP)} with \texttt{deepSIP}}
    \label{sec:deepSIP}
    
    We parameterize the spectra with \deepSIP \citep[][]{Stahl+2020_deepSIP}, a spectroscopic analysis framework capable of mapping preprocessed spectra (i.e., deredshifted, rescaled, and continuum-subtracted) to photometrically derived properties such as phase and light-curve shape\footnote{As described in the following paragraphs, we treat one of the output quantities from \deepSIP as a spectroscopic parameter. We compare photometrically derived counterparts to the \deepSIP output in Sec.~\ref{sec:photometric_dm15} and demonstrate that, through the neural network training with SN~Ia spectra and within the validated error size described in the following paragraph, \deepSIP is possibly capturing the ``secondary,'' spectroscopic features within its outputs. We discuss possible improvements to properly treat the spectroscopic components in Sec.~\ref{sec:discussions}.}. 
    \deepSIP accomplishes this by leveraging a specialized convolutional neural network (CNN) architecture trained on SN~Ia spectra from BSNIP, CFA, and CSP, with each spectrum coupled to \texttt{SNooPy} \citep{Burns_2011_SNooPy}-inferred quantities from the photometry of the underlying objects. Moreover, there are data augmentation operations performed to increase the size of the training data and make the model robust to varying signal-to-noise ratios (S/N), deredshifting errors, and other defects. One of the benefits to using a CNN as opposed to other approaches is that the model learns to encode complex features from inputs (i.e., processed flux densities) without the need for any manual feature identifications.
    
    For each input spectrum, \deepSIP provides five outputs: (i) a binary classification of whether the SN~Ia in the spectrum is at a rest-frame phase between $-10$ and 18\,d \emph{and} has a \dmfifteen \citep{Burns_2011_SNooPy} value between 0.85 and 1.55\,mag (a set of constraints which collectively define ``in-domain''), (ii, iii) the rest-frame phase of the spectrum and its uncertainty, respectively, and (iv, v) the \dmfifteen value and its uncertainty, respectively. 
    These five outputs are the concatenation of the results produced by three models which share a common neural architecture but are trained with different output objectives. In this framework, Model I produces (i) as a precursor to Models II and III, which produce (ii, iii) and (iv, v), respectively. Only those spectra for which (i) is classified as ``in-domain'' should have their continuous predictions for (ii--v) treated as actionable, and \cite{Stahl+2020_deepSIP} have demonstrated that when this is the case, the Phase and \dmfifteen value can be recovered with root-mean-square errors (RMSE) of just 1.00\,d and 0.068\,mag, respectively. Moreover, these metrics improve to 0.92\,d and 0.065\,mag (respectively) when the RMSE calculation is weighted inversely to the square of \deepSIP-predicted uncertainties, implying that, in the aggregate, when the model is more confident in its predictions, those predictions are more accurate.

    It is important to emphasize that the \dmfifteen value obtained by \deepSIP is, despite its unit of photometric magnitude(-decline), a spectroscopic quantity since the only input to \deepSIP is a spectrum of each SN. To avoid the confusion between photometric parameters ($m_B$, $x1$, $c$) measured by \texttt{SALT2} using the \texttt{Pantheon+} photometric data and this \deepSIP-derived parameter, we hereafter call this value SIP (``spectroscopically inferred parameter''). With the addition of SIP, we have three parameters ($c$, $x1$, and SIP) that can be used to characterize the luminosity of an SN~Ia (which is calculated from $m_B$ and the redshift).
    
    \subsection{Preprocessing}
    \label{sec:deepSIP-preprocessing}
    \begin{figure}[t]
        \centering
        \includegraphics[width=\linewidth]{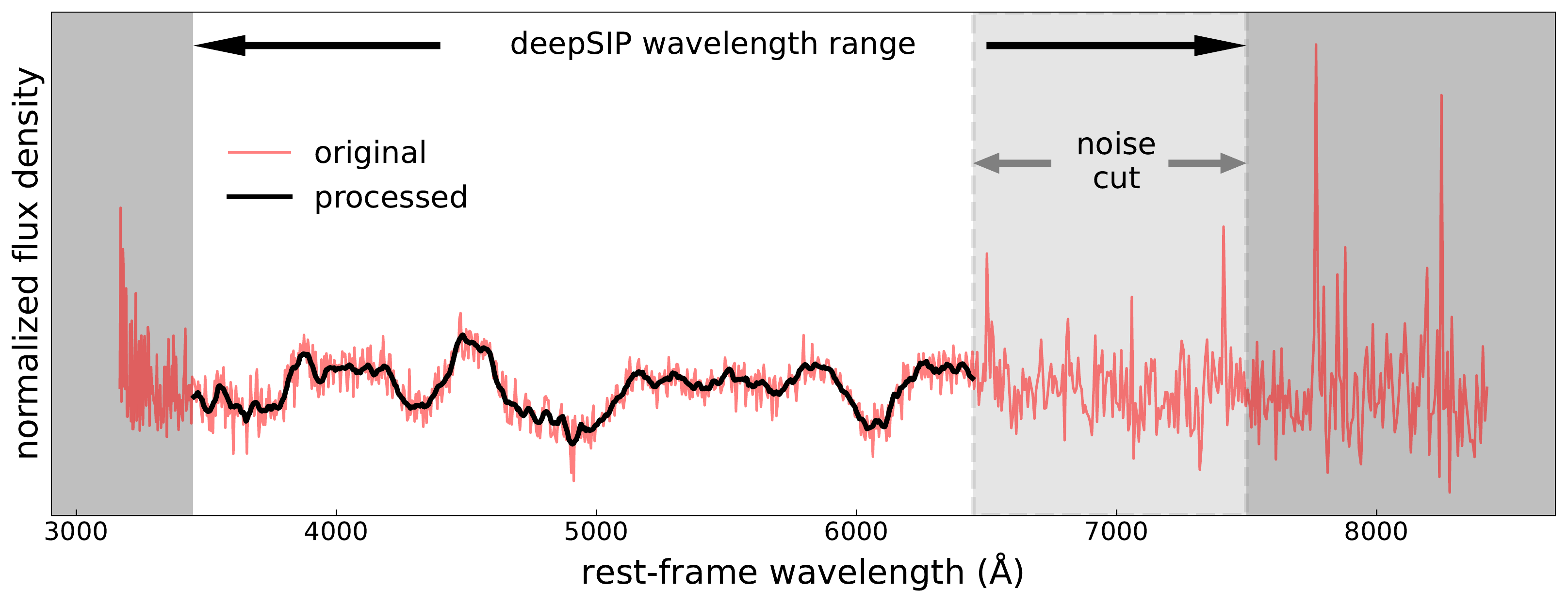}
        \caption{A demonstration of our preprocessing procedure on a spectrum of \texttt{SN2992} from \cite{Sako_2018_SDSS-SN-DR}. The thin, red line represents the raw spectral features after the continuum is removed. The black solid line represents the final output that is analyzed by \deepSIP after noise-cutting, smoothing, rebinning, and normalization are applied.}
        \label{fig:spec_processing_demo}
    \end{figure}
    
    Owing to the diverse sources in our compilation, the obtained spectra have a variety of forms; e.g., some come with quoted uncertainties, some contain only a narrow range of wavelength, some have very low S/N, and some are deredshifted.
    We account for these differences by (i) restoring data in the same format (simple array) of wavelength--flux, (ii) deredshifting the spectra when the supplied data are in the observer frame, and (iii) discarding the flux at wavelengths outside the 3450--7500\,\AA\ range over which \deepSIP operates.
    Since \deepSIP only analyzes the normalized spectral features rather than the flux value itself, the calibration is negligible and the unit of flux does not need to be converted.

    We then apply a series of preprocessing steps to encode spectra into the format expected by the neural-network-based \deepSIP. Most of this task follows the procedure described by \cite{Stahl+2020_deepSIP}. Fig.~\ref{fig:spec_processing_demo} demonstrates that some spectra have significant noise near the 7000\,\AA \  region, which falls within the range of wavelengths accepted by \deepSIP. When the whole \deepSIP range is used, this noise can lead to \deepSIP classifying the spectra as out-of-domain (see Sec.~\ref{sec:deepSIP} for the definition of ``in domain''). However, in many cases, the spectral noise in normalized spectra is limited to this small section of the continuum near 7000\,\AA\ (where continuum flux is small), and thus excising these contaminated data allows successful inference with \deepSIP.

    We validate such trimming by applying the same set of cuts to high-quality spectra that do not require it. In doing so, we find that certain combinations yield deviations (bias) in predicted \dmfifteen and Phase values. Our final set of cuts is chosen, therefore, such that the possible deviations are at most $\sim 1\sigma$ (see Fig.~\ref{fig:bias-analysis}; we describe this validation and bias analysis in Appendix~\ref{sec:dm15-cut-bias}).

    \subsection{Analysis with \deepSIP}
    \label{sec:deepSIP-analysis}

    \begin{figure}[t]
        \centering
        \includegraphics[width=\linewidth]{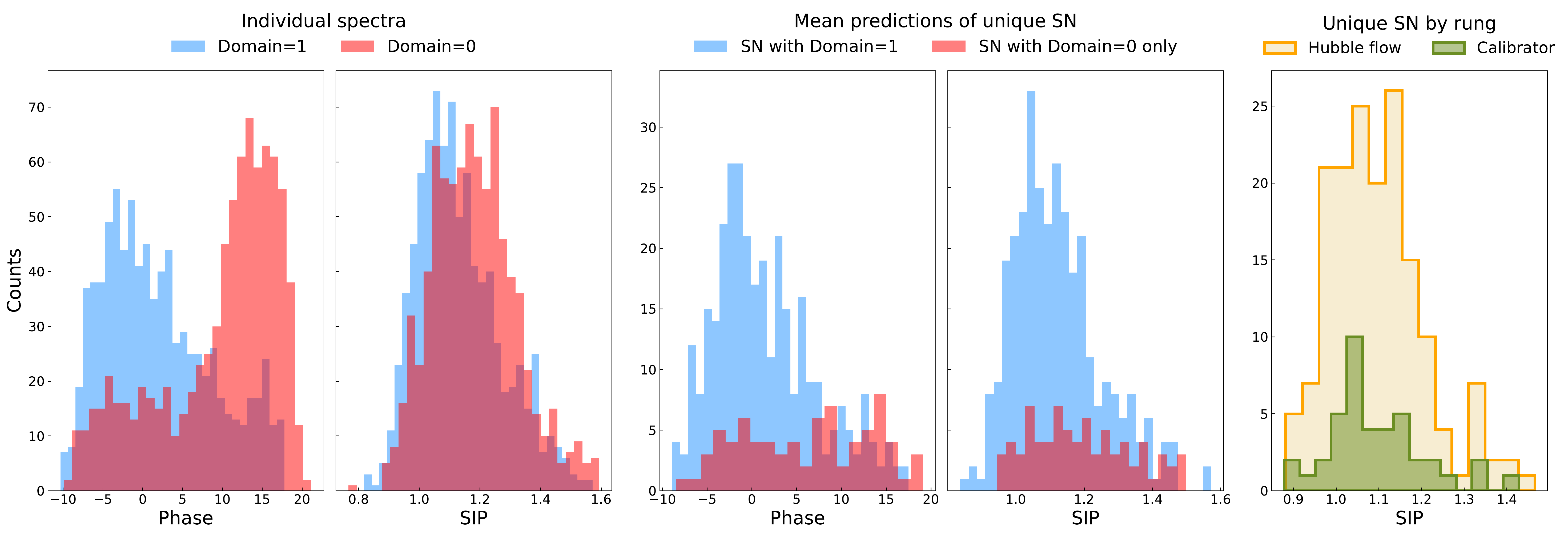}
        \caption{Distribution of estimated parameters for individual spectra (left panels) and the weighted mean values for each SN (middle and rightpanels). The left four panels compare the distributions by the \texttt{Domain} output of \deepSIP, and the right panel compares the distributions by rungs in the distance ladder.
        The blue histograms are for spectra (SNe) with \texttt{Domain=1}, which suggests that the spectra are in the domain of $-10\lesssim\text{phase}\lesssim 18$\,d and $0.85 \lesssim \text{SIP} \lesssim 1.55$\,mag as defined by \cite{Stahl+2020_deepSIP}. The out-of-domain spectra or SNe whose spectra are all out-of-domain are represented by the red histograms. The large offset between the peaks of the blue and red phase histograms (in comparison to the \dmfifteen histograms) suggests that the phase is the main culprit for the off-domain spectra. On the right panel, the distributions of SIP values are similar for both Cepheid-calibrated (dark green histogram) and Hubble flow (light orange histogram) SNe, suggesting that the populations are spectroscopically similar for both groups.}
        \label{fig:deepSIP-predictions}
    \end{figure}
    
    \begin{figure}
        \centering
        \includegraphics[width=\linewidth]{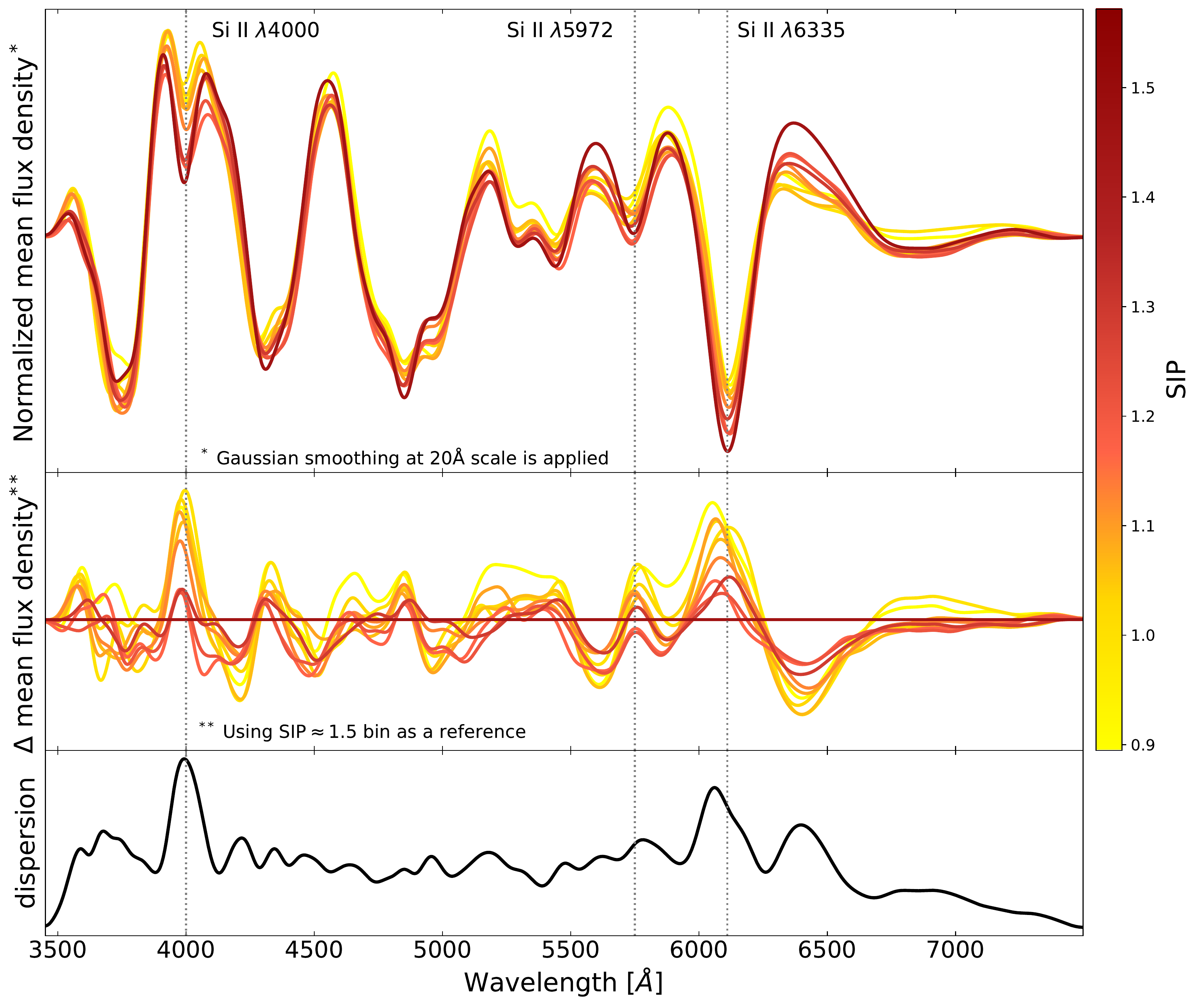}
        \caption{Intrinsic variation of SN~Ia spectra captured by SIP. Each colored line represents the mean (preprocessed) flux of SNe within equally-sized SIP bins. The relative differences, using SIP $\sim 1.5$ bin as a reference, are shown in the middle panel. The bottom panel represents the relative dispersion of spectra at each wavelength, in units of normalized, flattened (i.e., \deepSIP-ready) flux. Data in all panels are smoothed at a 20\,\AA\ scale to highlight the most significant variations. A selection of Si~II line locations is shown to qualitatively tie explosion physics to the SIP value.}
        \label{fig:spectral_diff}
    \end{figure}

    \begin{figure}
        \centering
        \includegraphics[width=\linewidth]{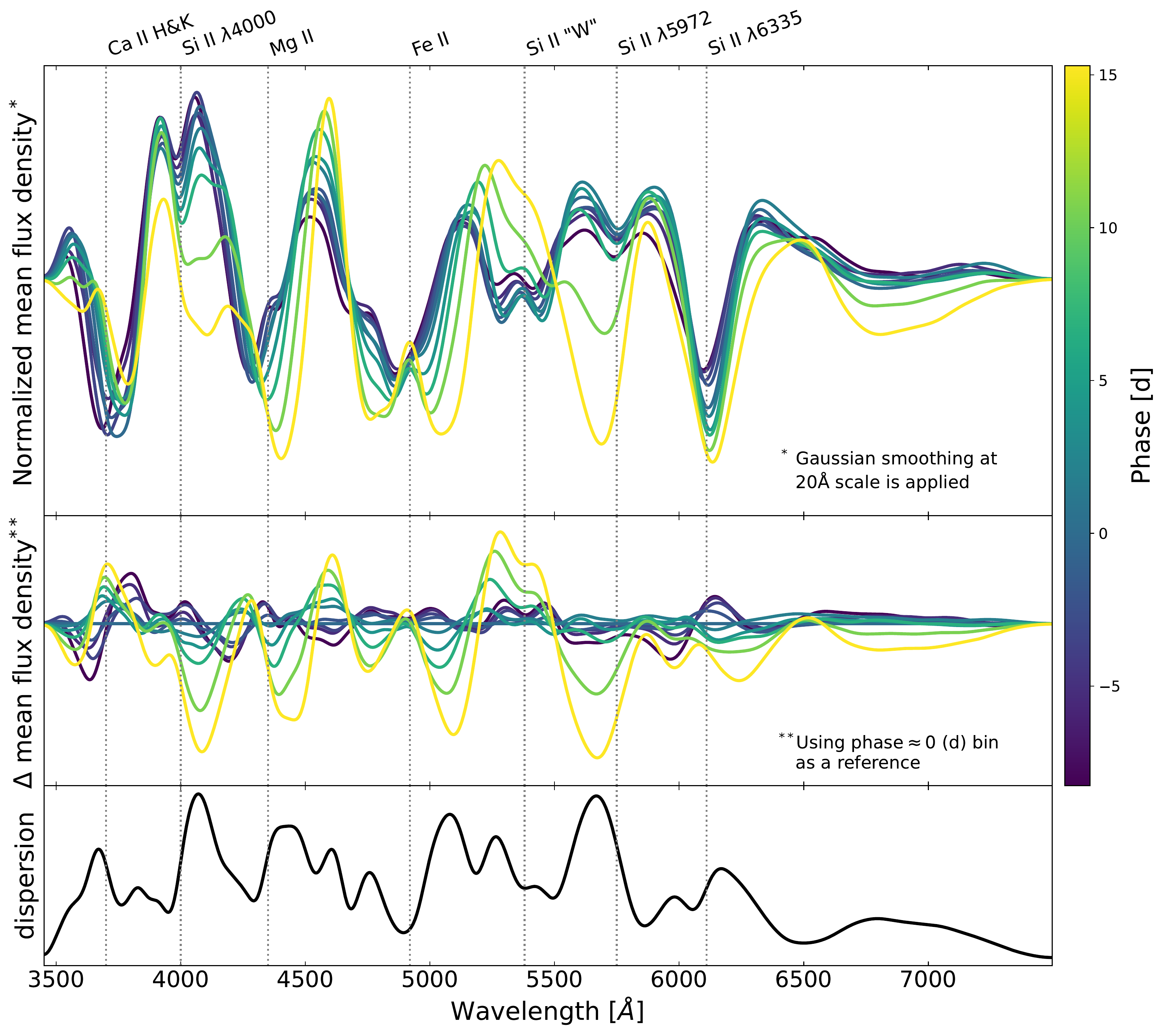}
        \caption{Temporal evolution of SN~Ia spectra along the \deepSIP-predicted phase values. All panels share the same format as in Fig.~\ref{fig:spectral_diff}. A selection of absorption features similar to those used by \cite{Stahl+2020_deepSIP,Stahl_2020_BSNIP} is shown to highlight the changes in ejecta velocities and optical depths over the span of $\sim 20$\,days. The wavelength-dependent dispersion is significantly different from Fig.~\ref{fig:spectral_diff}, indicating a clear separation between intrinsic and temporal variations.}
        \label{fig:spectral_diff_phase}
    \end{figure}

    We analyze all preprocessed spectra with \deepSIP without applying any photometry-based selection criteria (i.e., no temporal cuts). Our initial spectroscopic sample is therefore a mixture of ``in-domain" and ``out-of-domain" (e.g., late-time) spectra. 
    
    The resulting distributions of \deepSIP-predicted values are shown in Fig.~\ref{fig:deepSIP-predictions}. Of all spectra analyzed, $\sim 50$\% are classified as ``in-domain,'' and the resulting predicted phases are distributed around the time-of-maximum (i.e., $\sim 0$\,d). A significant fraction of predicted phases for ``out-of-domain" spectra are at $>10$\,d, peaking near the defined boundary of $\sim 15$\,d, and this indicates that most of the ''out-of-domain" results are due to the spectra being taken at late times.  The predicted SIP values share a similar distribution between ``in-domain" and ``out-of-domain" spectra, which further assures that the selection bias due to the domain of \deepSIP is minimal. 
    
    Furthermore, we compare the distributions of SIP values for the Cepheid-calibrated and Hubble flow SNe Ia (corresponding to the second and the third rung of the distance ladder, respectively) on the right panel of Fig.~\ref{fig:deepSIP-predictions}, analogous to Figure A2 of \citetalias{Riess_2022_SH0ES} in which similar comparisons were made for photometric quantities. The two distributions have the Kolmogorov–Smirnov statistic of $0.10$ with a $p$--value of $0.83$ ($\gg0.05$) when bootstrapped over the uncertainties associated with SIP; the distributions are indistinguishable, and it provides no evidence that the underlying physics of SN Ia which populate these rungs differ, at either the photometric or spectroscopic level.  The spectroscopic agreement between rungs, presented here for the first time, is a stronger test of SN similarity than prior photometric tests.

    The mean spectra of our sample, binned by SIP values, are visualized in Fig.~\ref{fig:spectral_diff}. Each bin contains equally numbered ($>80$) spectra that are classified as ``in-domain" by \deepSIP. The spectra in each bin span a broad range of (but mostly between $-5$ and 5\,d; see Fig.~\ref{fig:deepSIP-predictions}) phases, making the mean spectra most resemble those at the time of maximum brightness. The differences of the mean spectra along the SIP values thus represent the intrinsic variations of spectral features\footnote{Note the preprocessing applied to the spectra: the SIP values are not dependent on the color, luminosity, of other continuum properties.} captured by \deepSIP. The differences are most notable near Si~II lines, and such variations are continuous along the SIP values, consistent with previous studies \citep[e.g.,][]{Branch_2006,Blondin_2012}. The wavelength-dependent dispersion, shown in the bottom panel of Fig.~\ref{fig:spectral_diff} exhibits qualitative agreement with Fig.4 of \citetalias{Boone_2021_TwinsEmbeddinng_I}, though the effect of normalization applied during preprocessing (most notably \deepSIP's boundary conditions at 3500\,\AA and 7500\,\AA) makes the direct comparison difficult.

    The intrinsic variations highlighted in Fig.~\ref{fig:spectral_diff} show, as expected, a significantly different trend from the temporal variations displayed in Fig.~\ref{fig:spectral_diff_phase}, in which the mean spectra of ``in-domain" instances are shown for each equally-sized bin in phase space. A few key, well-known signatures of SN~Ia spectral evolution \citep[e.g.,][]{Filippenko_1997_SNspecReview,Branch_Wheeler_2017_SNtextbook} --- ejecta velocities slowing down at different rates for each element and evolving optical depths --- are visibly present. It is also worth noting the scale of variations along the phase space (Fig.~\ref{fig:spectral_diff_phase}) relative to the spectral features themselves, which is much larger compared to the variation scale along SIP space (Fig.~\ref{fig:spectral_diff}).

\section{Quantifying the scatter profile} 
    \label{sec:SP}

    With all relevant quantities prepared, we first define the ``\emph{Scatter Profile}'' ---  the (binned) dispersion of luminosity as a function of the pairwise similarity of independent observables. If our \texttt{SALT2} standardization is perfect or the observable is unrelated to luminosity, a flat scatter profile is expected (i.e., constant pairwise scatter). This scatter profile will be used to evaluate our new scatter model later in Sec.~\ref{sec:COV}.

    \subsection{Scatter size and similarity}
    \label{sec:SP-definition}

    For the evaluation of scatter, we first calculate the pairwise difference in standardized luminosity for objects $i$ and $j$ in the Hubble-flow sample (SNe-HF),
    \begin{align}
        \nonumber
        \Delta M_{B,ij} &= M_{B,j} - M_{B,i}\\\nonumber
                        &= \left[m_{B,j}-5\log cz_j\{\}\right] - \left[m_{B,i}-5\log cz_i\{\}\right]\, ,
    \end{align}
    for $i\ne j$, where the magnitude--redshift relation $M_B - 5\log H_0 = m_B - 5\log cz\{\} -25$ is used to calculate the individual standardized luminosity. The $\log cz\{\}$ term, together with $m_B$, measures an arbitrary expansion history (for the full formalism, see \citetalias{Riess_2022_SH0ES}).
    Our goal here is to characterize $\Delta M_{B}$, measured at a certain similarity level $P$, as this provides us with a metric describing the unmodelled information embedded in the data.

    We then calculate the difference in the parameters\footnote{It should be noted that, of the three parameters mentioned, only $\Delta m_{15}$ is the SIP (spectroscopically inferred parameter). We include $x_1$ and $c$ as additional characteristic parameters, as our scatter profile analysis is used to both (i) empirically resize the BS21 model, and (ii) embed the SIP kernel in the covariance matrix.}, i.e., $c$, $x_1$, and SIP between any pair: 
    \begin{equation}
        \delta_{\mathrm{SIP},ij} = \left|(\text{SIP})_i - (\text{SIP})_j\right|\, .
    \end{equation}
    
    For the calculated matrix $\delta$, the SIP rank between the $i$th and $j$th SN is defined as a percentile in $\delta_\text{param}$, which we denote as $P_{\delta,ij}$.
    We use only the lower-triangular elements of the matrices $\delta_{\mathrm{SIP},ij}$ and $\Delta M_{B,ij}$ to avoid double-counting.

    Finally, the SN~Ia luminosity scatter ($\mathcal{S}$) is measured as a standard deviation within equally-sized percentile bins:
    
    \begin{equation}\label{eq:std_scatter}
        \mathcal{S}(P_\text{bin}) = \sqrt{\frac{1}{2N_\text{pairs}}\sum_{(i,j)\in \text{bin}}\left(\Delta M_{B,ij}^2\right)}\, .
    \end{equation}
    The factor of $1/\sqrt{2}$ exists to account for taking the difference of luminosities in each pair rather than calculating the residual from the mean. In Eq.~\ref{eq:std_scatter}, $N_\text{pairs}$ denotes the number of pairs in each bin.
    
\begin{figure}
        \centering
        \includegraphics[width=\linewidth]{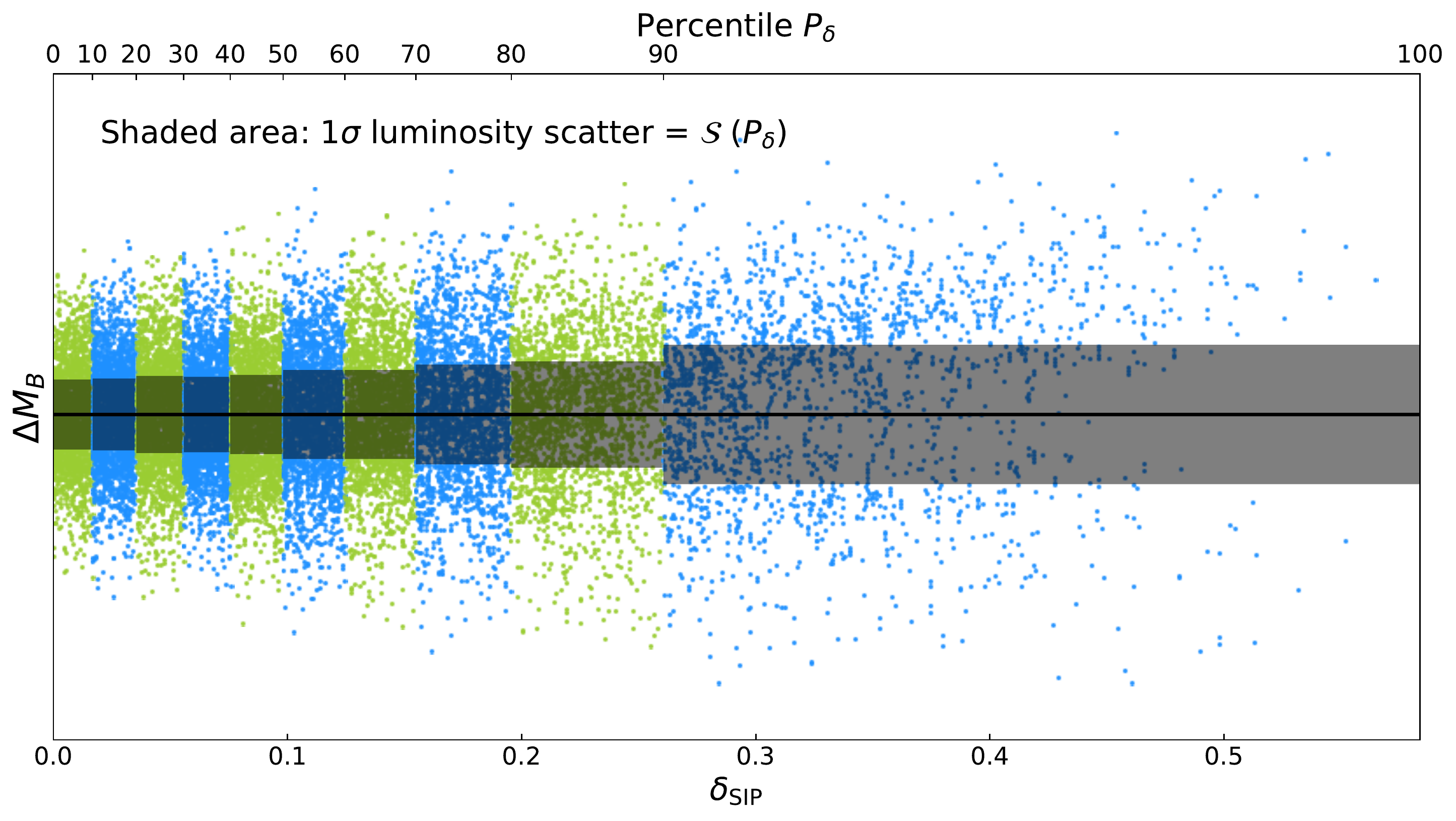}
        \caption{A conceptual demonstration of the pairwise scatter analysis presented in Sec.~\ref{sec:SP}. Each colored dot represents a pair of Hubble-flow SNe (see Sec.~\ref{sec:data}). The bottom horizontal axis represents the absolute difference in SIP values, and the top horizontal axis represents the corresponding percentile based on the distribution of $\Delta m_{15}$ values.}
        \label{fig:twinness_scater_demo}
    \end{figure}

    As a result, for each characteristic parameter ($c$, $x_1$, and SIP), we obtain three matrices ($M_{B,ij}$, $\delta_{ij}$, and $P_{\delta,ij}$), along with an array of intrinsic scatter ($\sigma_M$) for the given percentile-difference bins. We use the percentile $P_{\delta,ij}$ instead of the parameter distance ($\delta_{ij}$) to achieve an accurate measurement of the scatter with equally-sized bins. This process is visually demonstrated in Fig.~\ref{fig:twinness_scater_demo}.

    \subsection{Uncertainty}
    \label{sec:uncertainty}
    The methodology described above treats all observed quantities, including the luminosity $M_B$ and characteristic parameters ($c$, $x1$, and SIP), as uncertainty-free point values. 
    To account for parameter uncertianties, we resample the measurements many times (i.e., bootstrapping)
    to quantify uncertainties.
    In Fig.~\ref{fig:scatter_profile_c_x1} and Fig.~\ref{fig:scatter_profile_SIP}, which will be discussed further in Sec.~\ref{sec:SP_implication}, each thin line represents a result from a single sampling, and the thick line represents the mean across samples, which we treat as the overall result. This process is also visualized within the left panel of Fig.~\ref{fig:hyperparam_concept}.

    \subsection{Predicted scatter profile from covariance matrix}
    \label{sec:SP_prediction_from_cov}
    In addition to measuring the scatter of data, our analysis in Sec.~\ref{sec:COV} heavily relies on predicting (i.e., simulating) the scatter based on a given covariance matrix. In an ideal case, the predictions stemming from the covariance matrix should share the same size and structure as the scatter observed in any parameter space --- this is because the covariance matrix is fundamentally a description of the scatter in the data. For simulated data, we apply the same statistics as the observed data (Sec.~\ref{sec:SP-definition}) for comparison. In Fig.~\ref{fig:scatter_profile_c_x1} and Fig.~\ref{fig:scatter_profile_SIP}, the analysis results from these simulation-based data, evaluated in each sampled parameter space, are represented by thin pink lines. The thick pink line represents the mean of the simulated and sampled results.

\section{Scatter profile: results and implication}
    \label{sec:SP_implication}
\begin{figure}
        \centering
        \includegraphics[width=\linewidth]{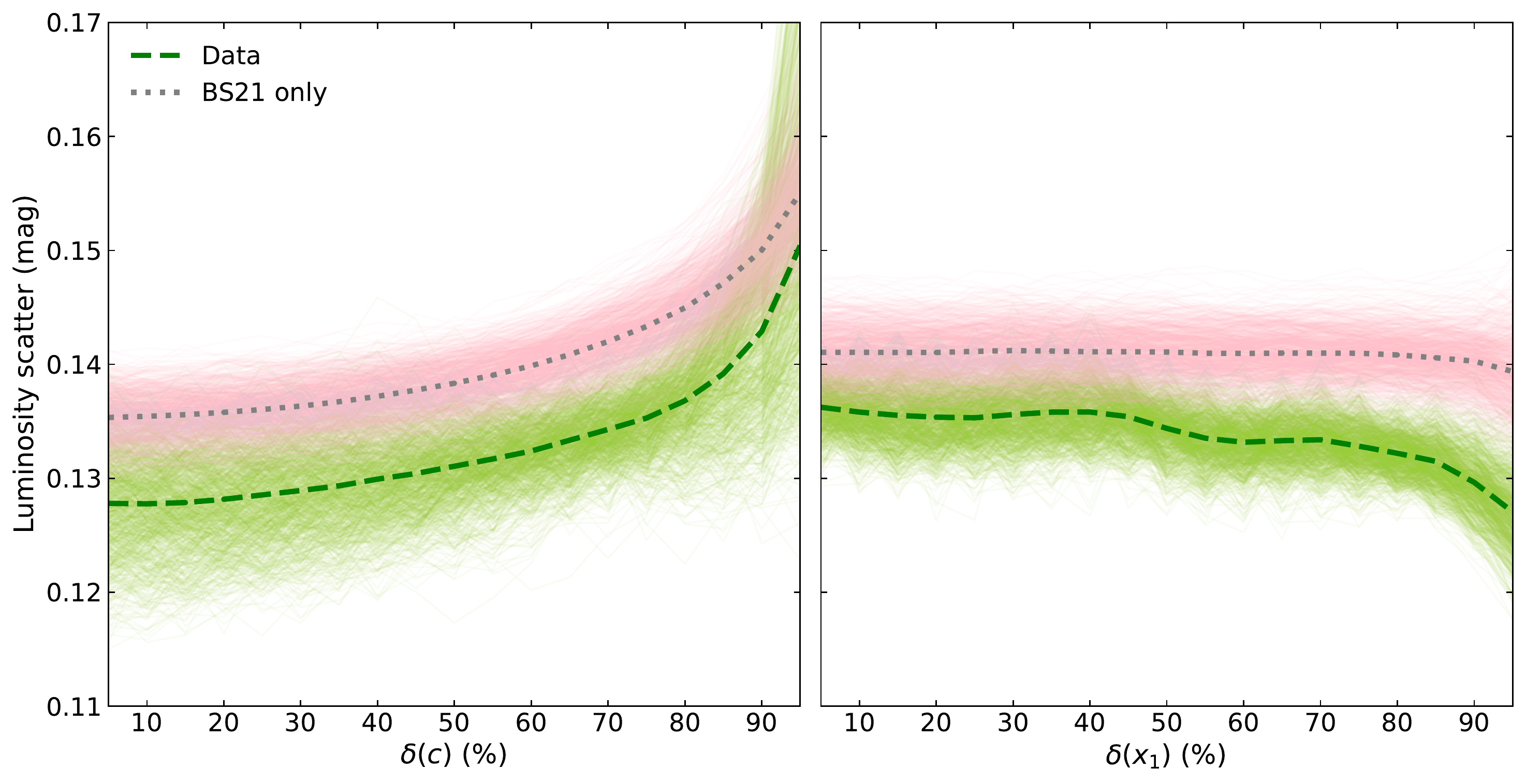}
        \caption{Scatter profiles measured with $c$ and $x_1$ similarities. The measured scatter is equivalent to the half-width of the shaded region in Fig.~\ref{fig:twinness_scater_demo}. As described in Sec.~\ref{sec:uncertainty}, thin lines represent sampled results, and thick dotted/dashed lines are the mean of the samples (i.e., the result of our measurements). Green (dashed) represents the observed data, and pink (dotted) are the simulated data based on the covariance matrix used in the SH0ES analysis (see Sec.~\ref{sec:SP_prediction_from_cov}). }
        \label{fig:scatter_profile_c_x1}
    \end{figure}

\begin{figure}
        \centering
        \includegraphics[width=\linewidth]{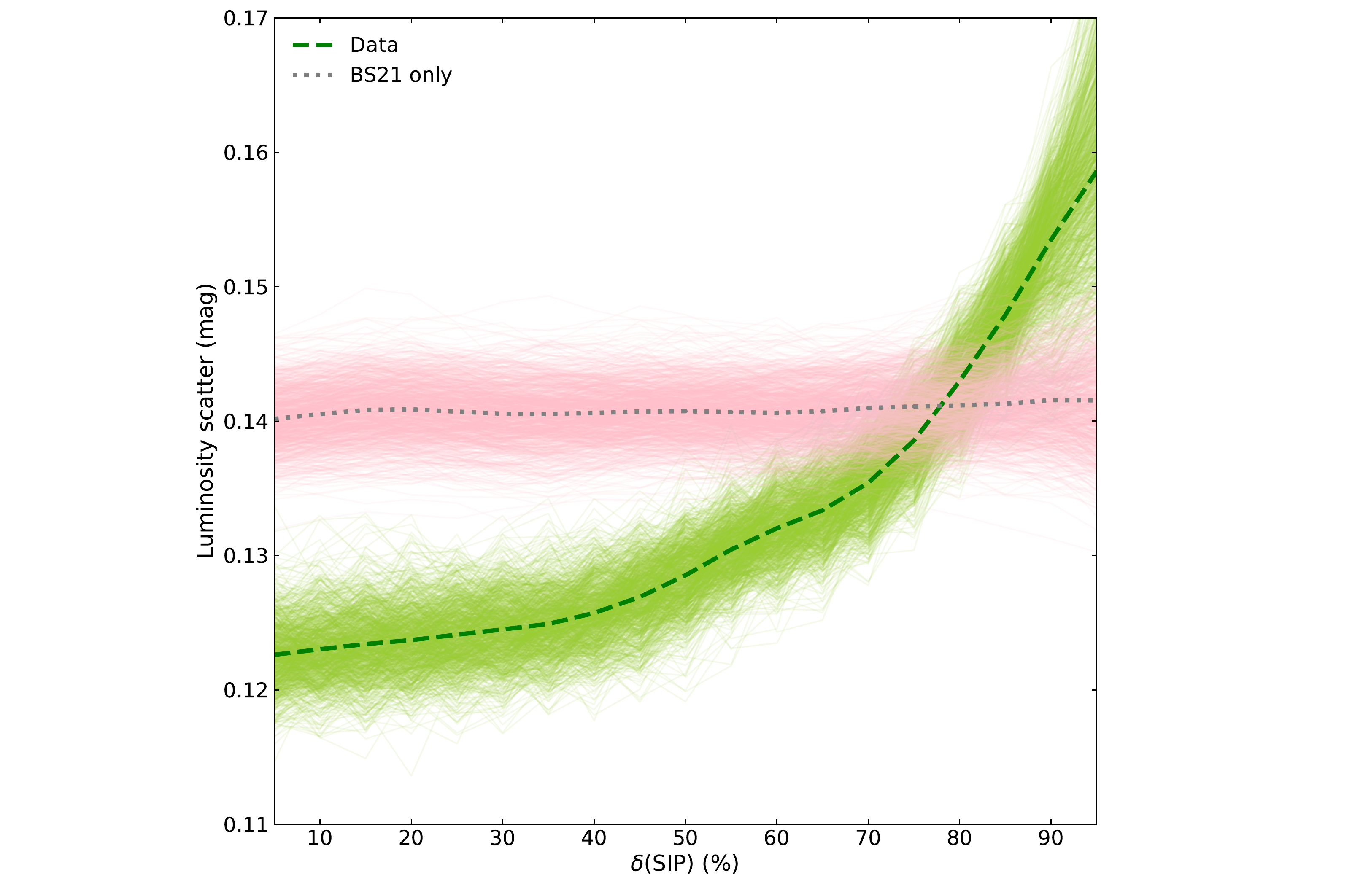}
        \caption{Scatter profiles measured with SIP similarities, shown in the same style as in Fig.~\ref{fig:scatter_profile_c_x1}. The relationship between luminosity scatter and SIP similarity constitutes a newly discovered trend, offering the possibility to effectively reduce the SN~Ia luminosity scatter to $\sim 0.123$\,mag.}
        \label{fig:scatter_profile_SIP}
    \end{figure}

    The measured scatter profile in the \texttt{SALT2} photometric parameters ($c$, $x_1$) and in \texttt{deepSIP}'s spectroscopic parameter (SIP) are shown in Fig.~\ref{fig:scatter_profile_c_x1} and Fig.~\ref{fig:scatter_profile_SIP}, respectively. Each observed scatter profile (green, dashed) is overlaid with a prediction of the scatter (pink, dotted) based on the BS21 covariance matrix (see Sec.~\ref{sec:SP_prediction_from_cov} for details). When the covariance matrix successfully depicts the features embedded in the data, the two (prediction and data) scatter profiles should agree. Both Fig.~\ref{fig:scatter_profile_c_x1} and Fig.~\ref{fig:scatter_profile_SIP} demonstrate the deviation of the current model from the data. Below we describe two major implications of the scatter profile measurement results.

    \subsection{Overestimated uncertainty} \label{sec:overestimated_uncertainty}
    The scatter profiles in the parameters used in the photometric model ($c$, $x_1$) 
    show good agreement with the model prediction: an upward trend of scatter in the $c$ (color) similarity rank matches the expectation \citep[see][]{Brout&Scolnic_2021}, and the nearly-flat scatter across $x_1$ space is also known and predicted.
    The $\sim 0.01$\,mag uniform offset in both spaces, however, indicates that the uncertainty size is overestimated compared to the scatter size of data. This may be a consequence of the tighter selection criteria ($|c|<0.15$, $|x|<2$) of \texttt{SH0ES} SNe~Ia within \texttt{Pantheon+} (see Appendix A.2 of \citetalias{Riess_2022_SH0ES}). The covariance matrix is generated by fitting all SNe~Ia in \texttt{Pantheon+} \citep[for parameter-dependent Hubble residuals, see][]{Brout_2022_P+Cosmo}, and the current model accounts for the mean scatter for the full range dataset.
    We will describe our method to account for this overestimated uncertainty in Sec.~\ref{sec:COV-gray}.

    \subsection{New trend in SIP-similarity} \label{sec:SIP_similarity_trend}
    In addition to the aforementioned offset, the clear trend in the newly added parameter space (SIP similarity; see Fig.~\ref{fig:scatter_profile_SIP}) indicates that there is a large amount of information embedded in the spectroscopic measurement that the current BS21 model does not take into account. Considering that this profile is a summary of pairwise comparisons, the trend --- that lower-rank pairs in SIP similarity have smaller scatter --- implies that \emph{spectroscopically similar SNe have similar luminosities after standardization}, which is consistent with numerous findings in the literature \cite[e.g.,][]{Fakhouri_2015_Twins,Boone_2021_TwinsEmbedding_II}.
    It is important to underscore that SIP is exclusively derived from spectroscopic information, and we can use the yielded values (SIP) as independent\footnote{If we were to use \texttt{SNooPy} to measure $\Delta m_{15}$ from photometric data, the measurement uncertainties would be common with those of the \texttt{SALT2} $x_1$, presenting obstacles in using it as an additional parameter.} measurements of SN properties (in addition to the existing \texttt{SALT2} parameters, $m_B$, $c$, $x_1$).

    The scatter size in SIP space varies from $\sim 0.123$\,mag for $\delta_\text{SIP}\sim 0$ to $\sim 0.159$\,mag for large $\delta_\text{SIP}$, rather than the mean of $\sim 0.14$\,mag. This indicates that an improved covariance model matching the trend shown by the data in Fig.~\ref{fig:scatter_profile_SIP} can reduce the effective luminosity scatter size to the observed minimum of $\sim 0.123$\,mag, subject to the availability of spectroscopic data in the calibrator sample. As mentioned in Sec.~\ref{sec:intro}, the luminosity scatter size accounts for 60--70\% of the current $H_0$ uncertainty, and thus this reduction of scatter has the leverage to yield a significant improvement in constraining power. Our method to construct an improved covariance model is discussed in Sec.~\ref{sec:COV-kernel}.

\section{Scatter Profile Modeling: updated covariance matrix}

    \begin{figure}
        \centering
        \includegraphics[width=\linewidth]{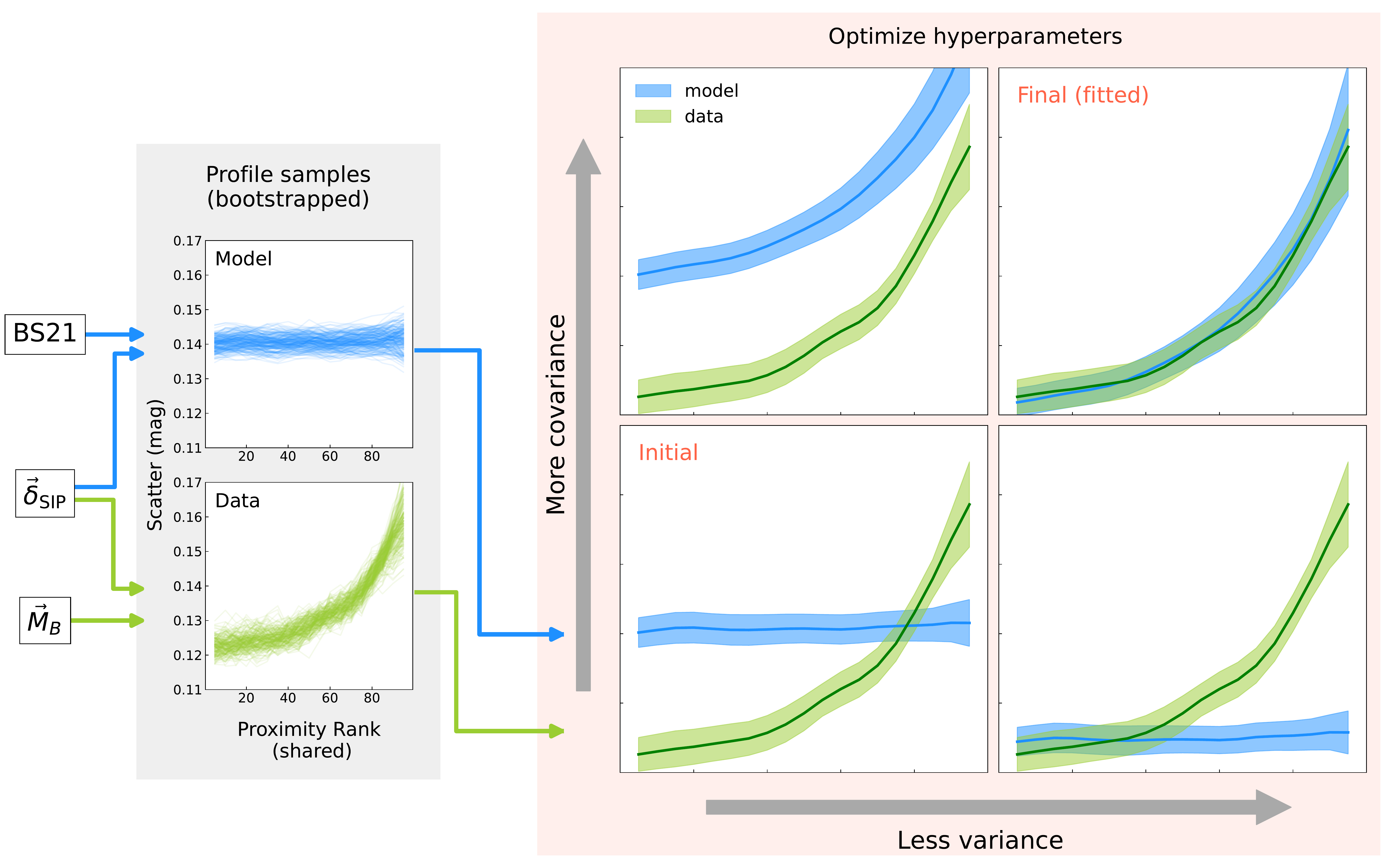}
        \caption{Visual concept of the process to optimize hyperparameters. From left to right, we (i) prepare observed quantities and the covariance matrix (Sec.~\ref{sec:data}, \ref{sec:deepSIP}), (ii) bootstrap the scatter profile measurement (Sec.~\ref{sec:uncertainty}, \ref{sec:SP-definition}, \ref{sec:SP_prediction_from_cov}; for both covariance model and data), and (iii) adjust the hyperparameters ($\Delta\sigma_\text{gray}$,$\sigma_\text{SIP}$,$l$) to fit the model (blue) to data (green).}
        \label{fig:hyperparam_concept}
    \end{figure}

    \label{sec:COV}
    The two primary findings described in Sec.~\ref{sec:SP_implication} (overestimated uncertainty and trend in SIP-similarity scatter profile) suggest that we can improve the current uncertainty modeling by (i) reducing the mean variance to match the data and (ii) parameterizing the covariance in SIP-similarity to model spectroscopic similarity.

    We modify the existing model by linearly combining a new covariance matrix with the existing one,
    since the existing covariance matrix for SNe~Ia includes a nontrivial information regarding cross-calibration, observational conditions \citep[][]{Scolnic_2021_P+data}, parameters' inter-dependencies, as well as the \texttt{BS21} scatter model that already accounts for the color-dependent scatter. We therefore aim to retain the successful components and only modify the covariance where needed. We assume that our new scatter model is sufficiently similar to the current one that BBC bias corrections --- which are sensitive,  at second order, to the variance properties of the data --- will not need to be recomputed.

    Following the observations described above, we invoke two models to calculate improved results. The first (Sec.~\ref{sec:COV-gray}) only accounts for the overestimated uncertainty, while the second  (Sec.~\ref{sec:COV-kernel}) accounts for the new trend with SIP. Since the second model requires controlling for the overestimated uncertainty simultaneously, it contains the first model and can therefore be considered as our fiducial analysis.
    The overall concept of our model, as well as the parameter fitting, is visualized in Fig.~\ref{fig:hyperparam_concept}.

    \subsection{Model 1: reduced gray scatter}
    \label{sec:COV-gray}
    The \texttt{BS21/P21} model is color-dependent, and within its formalism, it contains ``gray'' scatter due to the lower limit of the color-dependent effect. As it is not dependent on the color effect, gray scatter provides a variance-like behavior. Reducing this component can solve the overestimated uncertainty discussed in Sec.~\ref{sec:overestimated_uncertainty}.
    While the ``gray'' term behaves like variance (i.e., the diagonal component) in the covariance matrix, it is a physical property of individual SNe, and we need to account for the photometric duplicates \citep[same SN observed by multiple surveys; see][for details]{Scolnic_2021_P+data}. The ``gray'' term is therefore structured by the \emph{photometric duplicate matrix} based on the unique SN name \texttt{CID} in \texttt{Pantheon+} data:
    \begin{equation}
        D_{ij} = \begin{cases}
                    1 &  \text{if}\ \text{CID}_i =\text{CID}_j\\
                    0 &  \text{if}\ \text{CID}_i \ne \text{CID}_j\end{cases}\, .
    \end{equation}
    Denoting the size of the gray scatter to be removed as $\Delta \sigma_\text{gray}$, our new model is a linear combination of the original, \texttt{BS21} covariance matrix ($C_{\text{BS21}, ij}$) and the duplicate matrix $D_{ij}$:
    \begin{equation}
        \label{eq:model1}
        C_\text{model1} = C_{\text{BS21}, ij} - \Delta \sigma_\text{gray}^2 D_{ij}\, .
    \end{equation}
    Setting $\Delta \sigma_\text{gray}$ to zero reproduces the current BS21 model, and increasing the value of $\Delta \sigma_\text{gray}$ reduces the uncertainty. We optimize the value of $\Delta \sigma_\text{gray}$ using the method described in Sec.~\ref{sec:COV-fitting}.  We ensure that the optimized values of $\Delta \sigma_\text{gray}$ do not result in negative variance terms.  

    \subsection{Model 2: SIP-dependent covariance}
    \label{sec:COV-kernel}
\begin{figure}
        \centering
        \includegraphics[width=\linewidth]{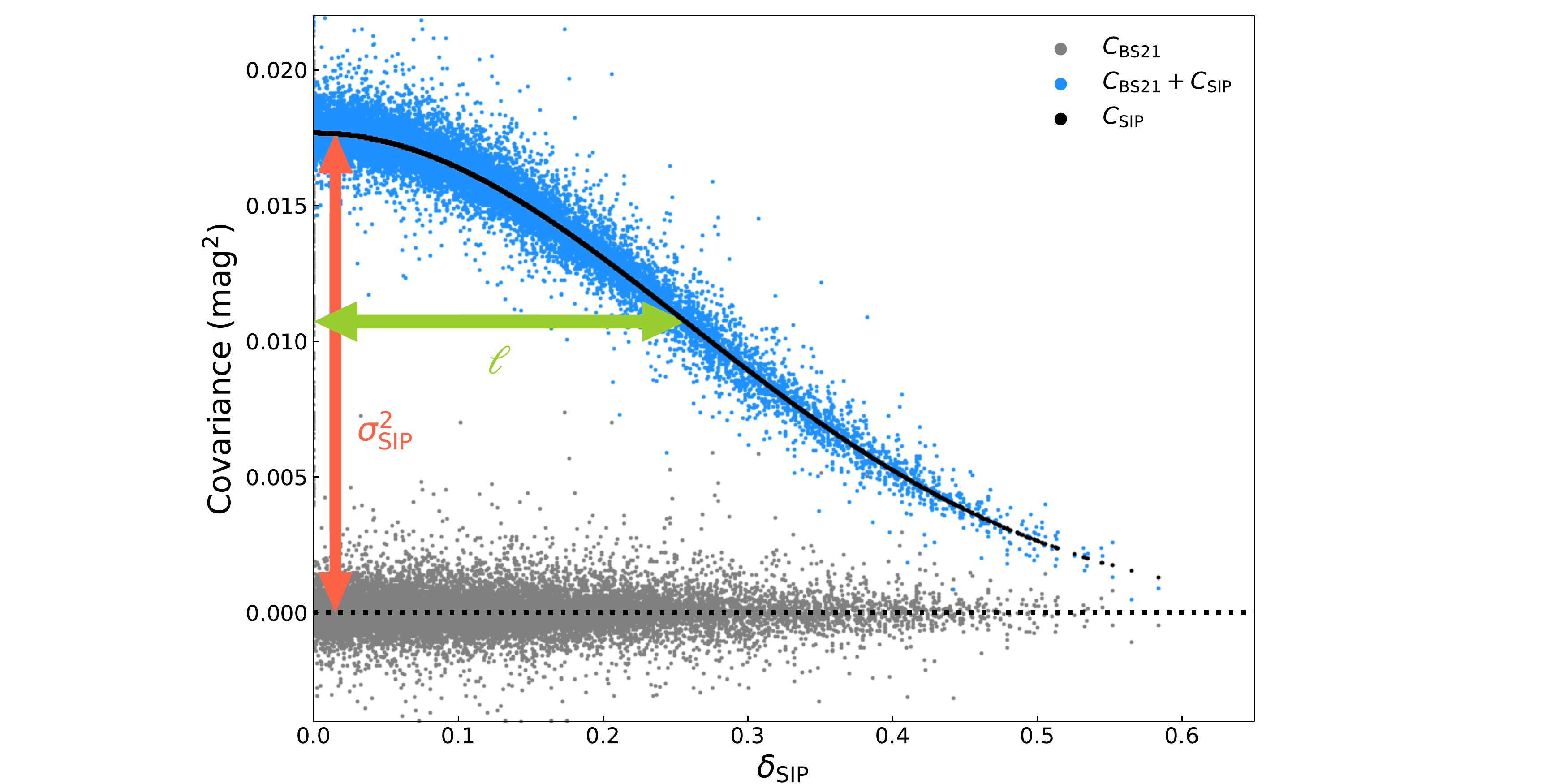}
        \caption{A comparison of covariance values (vertical axis) at each SIP distance (horizontal axis). Each scatter represents the assigned covariance value to each element of the matrix. The gray color shows the original \texttt{BS21/P21}-based covariance matrix, which has no structure in SIP space. The RBF kernel (black) is a pure function of SIP distance, which generates a Gaussian curve in the plotted space. These covariance matrices are then combined (Sec.~\ref{sec:COV-kernel}), which yields the values given in light-blue color. The shown RBF kernel represents the best-fit case of $(\sigma_\text{SIP}, \Delta\sigma_\text{gray},l) = (0.133,0.067,0.258)$ as determined in Sec.~\ref{sec:COV-fitting}. }
        \label{fig:covariance_scatter}
    \end{figure}

    To describe the $\delta_\text{SIP}$-dependent component of scatter, we use a radial basis function (RBF) kernel \citep[for a review of kernels, see Chapter 2 of][]{kernel_cookbook}. We denote this kernel (matrix) as $C_{\text{SIP}}$, whose elements are a function of the difference in parameter space relative to a scale $l$,
    \begin{equation}
    \label{eq:RBF-kernel}
        C_{\text{SIP},ij} = \exp\left(-\frac{\delta_\text{SIP}^2}{2l^2}\right)\, .
    \end{equation}
    We normalize this kernel by $\sigma_\text{SIP}^2$ as shown later in Eq.~\ref{eq:model2}. The exact size of the distance scale $l$ does not have a significant impact on our final results because the other free parameters described below can account for perturbations in its value. We optimize the parameter values from a grid in Sec.~\ref{sec:COV-fitting}.
    The covariance value solely from this RBF kernel is visualized as black dots in Fig.~\ref{fig:covariance_scatter}.

     The BS21/P21 model compensates for ignorance of the luminosity effect associated with SIP by increasing the gray scatter. With better understanding of the luminosity scatter, the unexplained gray scatter is expected to decrease. To account for this effect, we need to simultaneously reduce the variance similarly to model 1. Our final model is therefore a linear combination of $C_\text{BS21}$, $D$, and $C_\text{SIP}$:
    \begin{equation}
        \label{eq:model2}
        C_{\text{model2},ij} = C_{\text{BS21},ij} - \Delta \sigma_\text{gray}^2 D_{ij} + \sigma_\text{SIP}^2 C_{\text{SIP},ij}\, ,
    \end{equation}
    where the size of the RBF kernel is represented by $\sigma_\text{SIP}$. As $\sigma_\text{SIP}$ is increased (under optimal sizes of $\Delta \sigma_\text{gray}$), the estimated scatter between pairs with small $\delta_\text{SIP}$ is reduced (as they have large covariance), and the upward trend observed in the $\delta_\text{SIP}$ scatter profile (Sec.~\ref{sec:SIP_similarity_trend}) is reproduced. The effects of $\Delta \sigma_\text{gray}$ and $\sigma_\text{SIP}$ are conceptually visualized in Fig.~\ref{fig:hyperparam_concept}.

    \subsection{Evaluation and fitting} \label{sec:COV-fitting}

    \begin{figure}
        \centering
        \includegraphics[width=\linewidth]{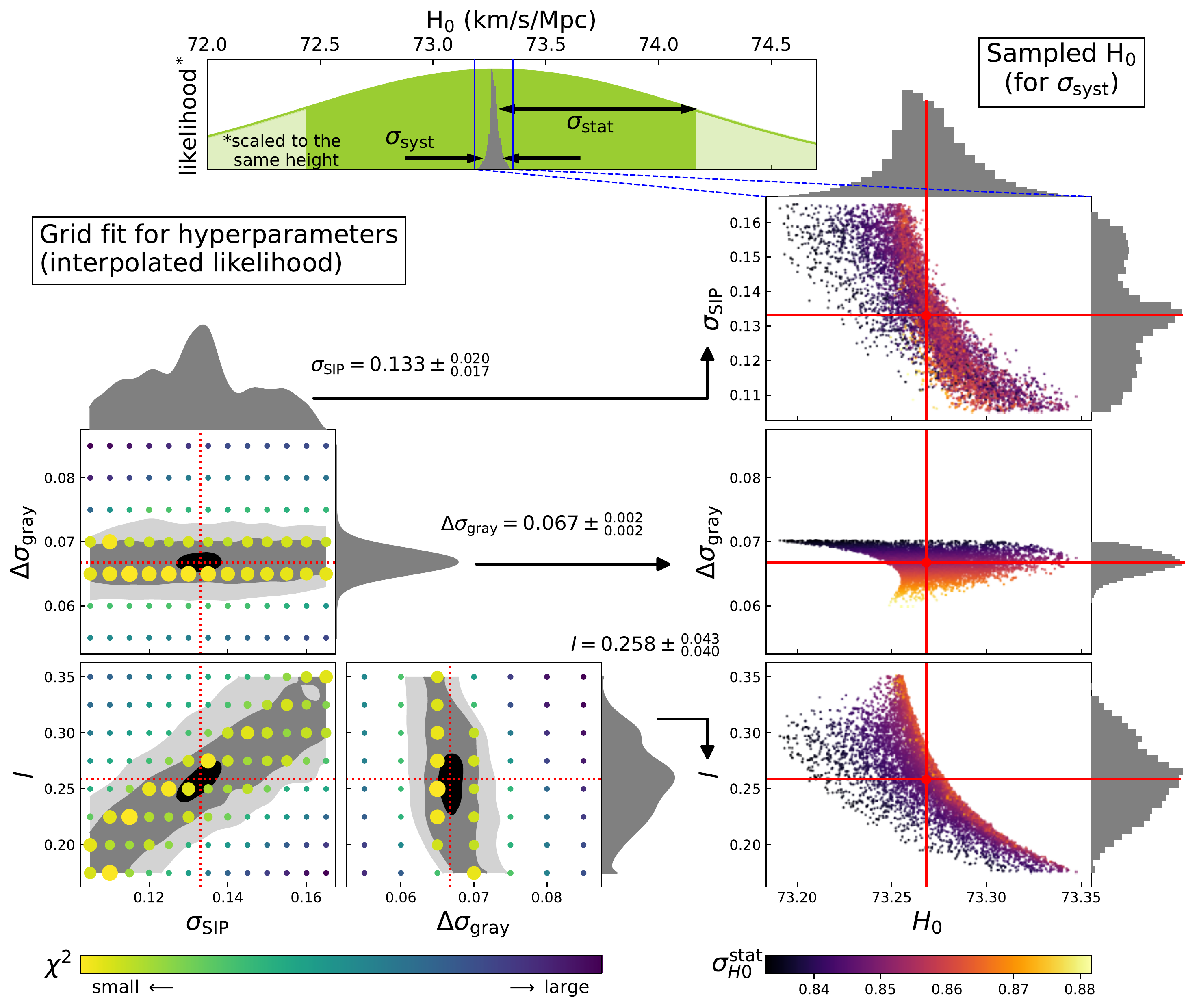}
        \caption{A visualization of our grid fit process (left) and \Hnaught sampling to estimate systematic uncertainty based on the grid fit results (right). \textbf{Left}: Each colored dot represents the gridpoint at which Fig.~\ref{fig:scatter_profile_fitted}-like scatter profiles are calculated for both data and our model with chosen hyperparameters (see each individual model-data comparison on the right panel of Fig.~\ref{fig:hyperparam_concept}). The colored regions (black, gray, and light gray) represent the interpolated confidence intervals at the 1$\sigma$, 2$\sigma$, and 3$\sigma$ level, respectively. \textbf{Right}: distributions of sampled hyperparameters, based on the likelihood determined (left panel), and the corresponding $H_0$ values. The top panel visualizes that the resulting $H_0$ distribution is significantly smaller than the values of uncertainty derived by solving the distance-ladder matrices (size of the shaded region).}
        \label{fig:grid+syst}
    \end{figure}

    \begin{figure}
        \centering
        \includegraphics[width=\linewidth]{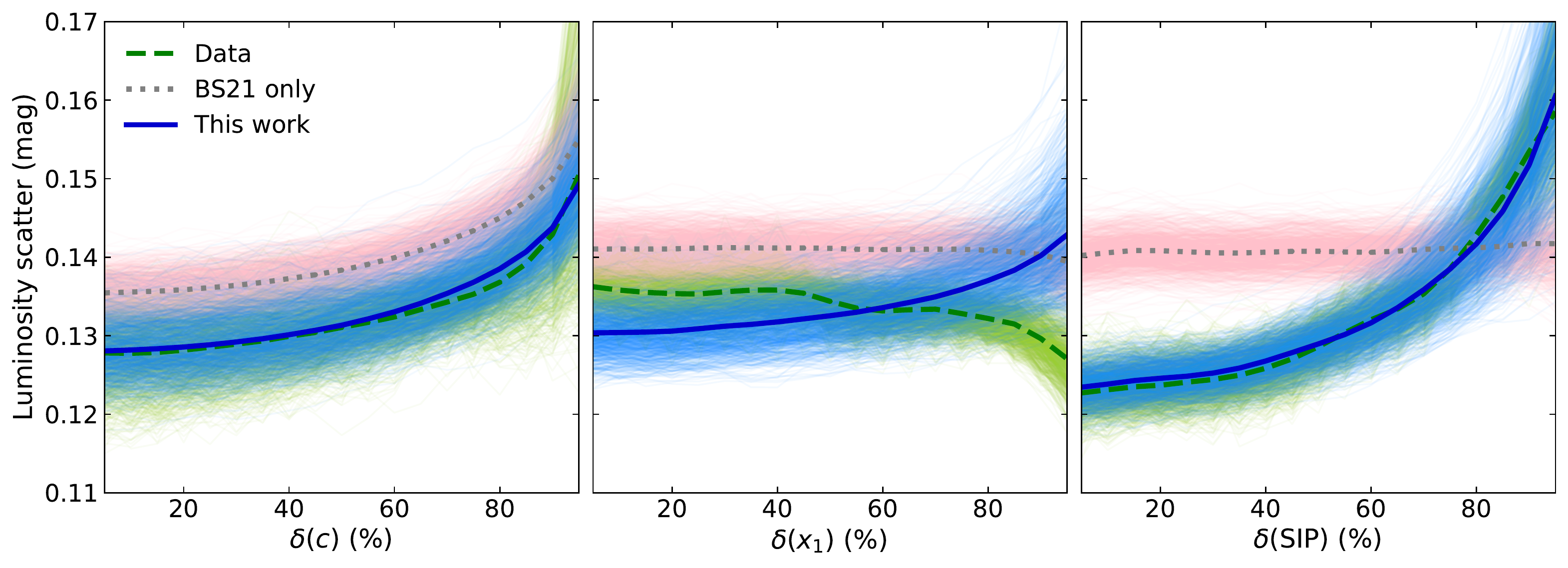}
        \caption{Comparison between observed scatter profiles and our covariance model. Individual samples and the mean of the best-fit model are shown in blue color and the solid thick blue line, respectively.  Green (dashed) represents the observed data, and pink (dotted) are the simulated data based on the covariance matrix used in the SH0ES analysis}
        \label{fig:scatter_profile_fitted}
    \end{figure}

    We identify optimal model parameters by minimizing the deviation between the model-predicted scatter profile and the scatter profile measured from Hubble flow SNe data\footnote{We use Hubble flow SNe only to make the measurement calibration-free and independent of Cepheid measurements.}. Following the discussion of the uncertainty in Sec.~\ref{sec:uncertainty}, model- and data- scatter profiles are both constructed over a large number ($\sim 1000$) of samples drawn from the observed parameters\footnote{The number of samples is chosen so that the resulting $\chi^2$ value converges to within $\sim 5$\% of the truth value.}
    \begin{eqnarray*}
        c \ &\sim& \ \mathcal{N}(c^\text{SALT2},\sigma_c^\text{SALT2})\\
        x_1 \ &\sim& \ \mathcal{N}(x_1^\text{SALT2}, \sigma_{x1}^\text{SALT2})\\
        \text{SIP} \ &\sim& \ \mathcal{N}(\text{SIP}^\text{deepSIP}, \sigma_\text{SIP}^\text{deepSIP})\ .
    \end{eqnarray*}
    The resulting $\sim 1000$ scatter profiles ($\mathcal{S}$; see the left panel of Fig.~\ref{fig:hyperparam_concept}) in each space are summarized into the mean and the standard deviation ($\mathcal{\overline{S}}$, $\sigma_\mathcal{S}$; see the middle panel of Fig.~\ref{fig:hyperparam_concept}), which are then used to evaluate the $\chi^2$ values over the percentile bins ($P$) and the three parameters ($i$):
    \begin{equation}
        \label{eq:chi_square}
        \chi^2 (\sigma_\text{SIP},\Delta \sigma_\text{gray},l) 
         = \sum_{i = \{c,x_1,\text{SIP}\}} \sum_P
            \frac{\left[\ \mathcal{\overline{S_\text{data}}}(P) - \mathcal{\overline{S_\text{model}}}(P)\ \right]_i^2}
            {\left[\sigma_{\mathcal{S}}^\text{data}(P)\right]_i^2 + \left[\sigma_{\mathcal{S}}^\text{model}(P)\right]_i^2}\, .
    \end{equation}
    While neither Model 1 nor Model 2 parametrize the covariance matrix as a function of \texttt{SALT2} parameters ($m_B$, $c$, and $x_1$) as inputs, we include the scatter profiles in those spaces as constraints so that we do not break an existing, successful covariance in those spaces: this is required owing to the partial correlation between SIP and $x_1$, and we further discuss this effect in Sec.~\ref{sec:discussions}.

    We perform the optimization by grid evaluation in $\Delta \sigma_\text{gray}$-space for Model 1 and ($\sigma_\text{SIP}$, $\Delta \sigma_\text{gray}$, $l$)-space for Model 2. The left panel of Fig.~\ref{fig:grid+syst} shows the slice of the grid at $l=0.2$. The $\chi^2$ value at each grid point is evaluated by Eq.~\ref{eq:chi_square}. The $\chi^2$ grid is then processed with cubic interpolation to determine the best-fit value and to obtain the likelihood contour (shown as gray confidence intervals in Fig.~\ref{fig:grid+syst}). The likelihood contour is used to evaluate the systematic error in Sec.~\ref{sec:distladder}.
    Our best fit result is shown in Fig.~\ref{fig:scatter_profile_fitted}. The best-fit parameters are $(\sigma_\text{SIP},\Delta \sigma_\text{gray},l) = (0.133\pm0.02\, \text{mag},\ 0.067\pm0.02\, \text{mag},\ 0.258\pm0.042\, \text{mag})$. Using these parameters, our new covariance model estimates the expected RMS differences of any pair of SNe, per each SN, (i.e., ordinate-axes values of Fig.~\ref{fig:scatter_profile_c_x1} and \ref{fig:scatter_profile_SIP}) as the following:
    \begin{eqnarray}\nonumber
        \langle\frac{1}{\sqrt{2}}\Delta M_B\rangle_{ij}^2 &=& \frac{1}{2}\left(C_{ii}^2 + C_{jj}^2 - 2C_{ij}^2\right)\\ \nonumber
        &=& 
        \frac{1}{2}\left[C_{\text{BS21},ii}^2 - (\Delta \sigma_\text{gray})^2 + \sigma_\text{SIP}^2\right] + 
        \frac{1}{2}\left[C_{\text{BS21},ii}^2 - (\Delta \sigma_\text{gray})^2 + \sigma_\text{SIP}^2\right] - 
        \left[C_{\text{BS21},ij}^2 + \sigma_\text{SIP}^2C_{\text{SIP},ij}^2\right]\\
        &=& \langle\frac{1}{\sqrt{2}}\Delta M_B\rangle_{\text{BS21},ij}^2 - 0.067^2 + 
        0.133^2\left[1 - e^{-\delta^2_{\text{SIP},ij}/(2\cdot0.258^2)}\right]\ .
    \end{eqnarray}

\section{The Hubble Constant} \label{sec:distladder}
    \begin{table}
        \centering
        \begin{tabular}{lcccccc}
            \hline
             Model & $\Delta \sigma_\text{gray}$ & $\sigma_\text{SIP}$ & $l$ & Calibration & $H_0$ & $\sigma_{H0}$ \\
             \hline
             BS21 & - & - & - & SH0ES 2022 & 73.04 & 1.01 \\
             BS21 & - & - & - & ClusterCeph & 73.04 & 0.97 \\
             ``$\Delta$gray'' (model 1) & 0.048 $\pm$ 0.008& - & - & ClusterCeph & 73.14 & 0.91 \\
            \textbf{``SIP''}  (model 2) & 0.067 $\pm$ 0.002 & 0.133 $\pm$ 0.020 & 0.258 $\pm$ 0.042 & ClusterCeph & \textbf{73.29} & \textbf{0.85} \\
             \hline
        \end{tabular}
        \caption{Results of distance-ladder evaluation with different covariance models. Previous values of the Hubble constant, from \citetalias{Riess_2022_SH0ES} (SH0ES 2022) and \citetalias{SH0ES_ClusterCeph} (ClusterCeph), are shown for reference. The values of $\sigma_\text{H0}$ include uncertainty directly evaluated from the distance ladder, and do not include the systematic uncertainty due to variants of fits (see \citetalias{Riess_2022_SH0ES}).}
        \label{tab:H0_results}
    \end{table}
    
    With the optimal hyperparameters determined for both Model 1 and Model 2 that best describe the scatter profile with Hubble flow SNe, we apply the model to all SNe (Cepheid calibrated and Hubble flow SNe; see Sec.~\ref{sec:data-cosmologydata} for details) to construct the full SN covariance matrix.
    SIP values are unavailable for 72 out of 277 Hubble flow SNe (missing spectra or all spectra ``out-of-domain"), and the variance or covariance values in the covariance matrix those SNe are kept unchanged -- that is, we do not benefit from correcting the overestimated uncertainties or accounting for the spectral similarities for them.
    We then evaluate the distance ladder using the updated SN covariance matrices, and employ the updated cluster Cepheid calibration of SNe (\citetalias{SH0ES_ClusterCeph}), replacing the SN components of the full covariance matrix with the SN covariance matrix of the corresponding model.

    Our results, as well as previous values from \citetalias{Riess_2022_SH0ES} and \citetalias{SH0ES_ClusterCeph}, are shown in Table~\ref{tab:H0_results}.
    Using the best-fit parameters for Model 2 (SIP-dependent model), we evaluate the distance ladder and obtain the local $H_0$,
    \begin{equation}
        \label{eq:H0}
        H_{0,\, \text{baseline}} = 73.29 \pm 0.85\ {\rm km}\,{\rm s}^{-1}\,{\rm Mpc}^{-1}\, .
    \end{equation}
    This is a $\sim 16\%$ improvement in uncertainty size compared to the previous \textit{baseline} result in \citetalias{Riess_2022_SH0ES} and $\sim 7\%$ improvement after the correction of overestimated uncertainty (Model 1, $\Delta$gray).

    In addition to the uncertainty presented above, which is solely based on a single covariance matrix (i.e., single set of hyperparameters), we evaluate the systematic uncertainty due to the choice of hyperparameters. We first draw samples of hyperparameters ($\sigma_{\text{gray},i}$, $\sigma_{\text{SIP},i}$, and $l_i$) weighted by the interpolated likelihood (based on the $\chi^2$ grid; see Sec.~\ref{sec:COV-fitting}). The resulting samples of hyperparameters represent the posterior of the scatter profile fitting. The $H_0$ values associated with the covariance matrix generated by each set of hyperparameters are calculated, and we take the size of the resulting distribution $\{H_{0,i}\}$ as the systematic error due to SIP scatter profile modeling. As shown in Fig.~\ref{fig:grid+syst}, the variation of $H_0$ value is almost negligible: we find the standard deviation $\sigma(\{H_{0,i}\}) = 0.023$, 16th--50th percentile separation $\Delta H_0^{50-16} = 0.017$, and 50th--84th percentile separation $\Delta H_0^{84-50} = 0.023$ in units of km\,s$^{-1}$\,Mpc$^{-1}$. We take the largest of those, $\sigma_\text{SIP--syst}= 0.023$ km\,s$^{-1}$\,Mpc$^{-1}$, as the most conservative estimation of the systematic uncertainty in our results. Combined with the baseline uncertainty ($\sigma_{H0} = \sqrt{\sigma_\text{baseline}^2 + \sigma_\text{SIP--syst}^2}$), our baseline measurement is
    \begin{equation}
        \label{eq:H0_baseline}
        H_{0,\, \text{baseline + SIP--syst}} = 73.29 \pm 0.85 \ {\rm km}\,{\rm s}^{-1}\,{\rm Mpc}^{-1}\ .
    \end{equation}

    \begin{figure}[t]
        \centering
        \includegraphics[width=\linewidth]{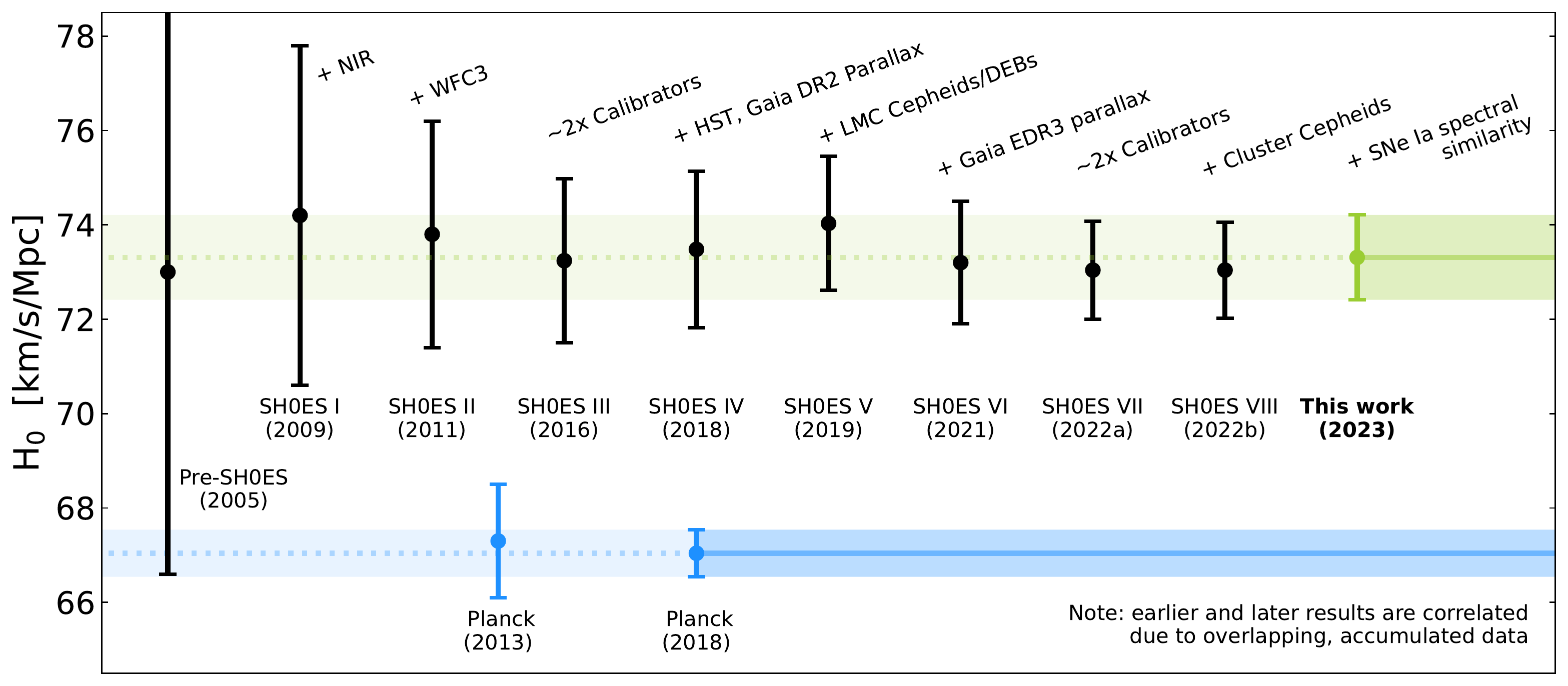}
        \caption{Progression of SH0ES results. Each iteration \cite{Riess_2005_Pre_SH0ES,Riess_2009a_SH0ES1a, Riess_2009b_SH0ES1b,Riess_2011_SH0ES2,Riess_2016_SH0ES3,Riess_2018a_SH0ES4_HST,Riess_2018b_SH0ES4_DR2, Riess_2019_SH0ES5_LMC, Riess_2021_SH0ES6_EDR3, Riess_2022_SH0ES, SH0ES_ClusterCeph} has implemented additional methods (as annotated in the figure) to tighten the local distance ladder, and this work provides the newest update using SN~Ia spectral similarity. The CMB measurement of \Hnaught by {\it Planck} \cite{Planck_2013,Planck_2018} is shown to visualize the Hubble tension, which has not been solved by any of the observational/methodological updates and improvements on the local distance ladder. For both SH0ES and {\it Planck} results, the most recent values and their uncertainties are shown as horizontal lines and colored area for visual aid.}
        \label{fig:SH0ES_results}
    \end{figure}    
    Finally, we also consider the additional systematic uncertainty that our previous analysis (\citetalias{Riess_2022_SH0ES}) included: an extra $\sim 0.3$\,km\,s$^{-1}$\,Mpc$^{-1}$ uncertainty was measured over 67 variants of the SH0ES distance ladder, and this provides a final, conservative adjustment to account for possible biases of (from) the baseline set due to analysis methods, such as selections of data at each rung, the inclusion of tip of the red-giant branch (TRGB), selection of SN scatter model, peculiar-velocity correction, and exclusion of near-infrared observations. Following \citetalias{Riess_2022_SH0ES}, we add this additional systematic uncertainty ($\sigma_\text{variants}$) in quadrature ($\sigma_{H0} = \sqrt{\sigma_\text{baseline}^2 + \sigma_\text{SIP--syst}^2 + \sigma_\text{variants}^2}$). 
    We present our fiducial measurement of the Hubble constant,
    \begin{equation}
        \label{eq:H0_final}
        \mathbf{H_0 = 73.29 \pm 0.90 \ \ {\rm km}\,{\rm s}^{-1}\,{\rm Mpc}^{-1}\ .}
    \end{equation}
    Our result provides the most precise value of \Hnaught measured by the local distance ladder. The increased, $\sim5.7\sigma$ discrepancy between the local \Hnaught and the $\Lambda$CDM calibrated by \citetalias{Planck_2018} suggests that the spectral variation of standardized SNe~Ia does not explain the ``Hubble tension,'' similarly to all improvements employed previously, as shown in Fig.~\ref{fig:SH0ES_results}.

\section{Discussion} \label{sec:discussions}

    \subsection{What is \textit{SIP}? Comparison to the photometric estimation of light-curve shapes} \label{sec:photometric_dm15}
    
\begin{figure}[t]
        \centering
        \includegraphics[width=\linewidth]{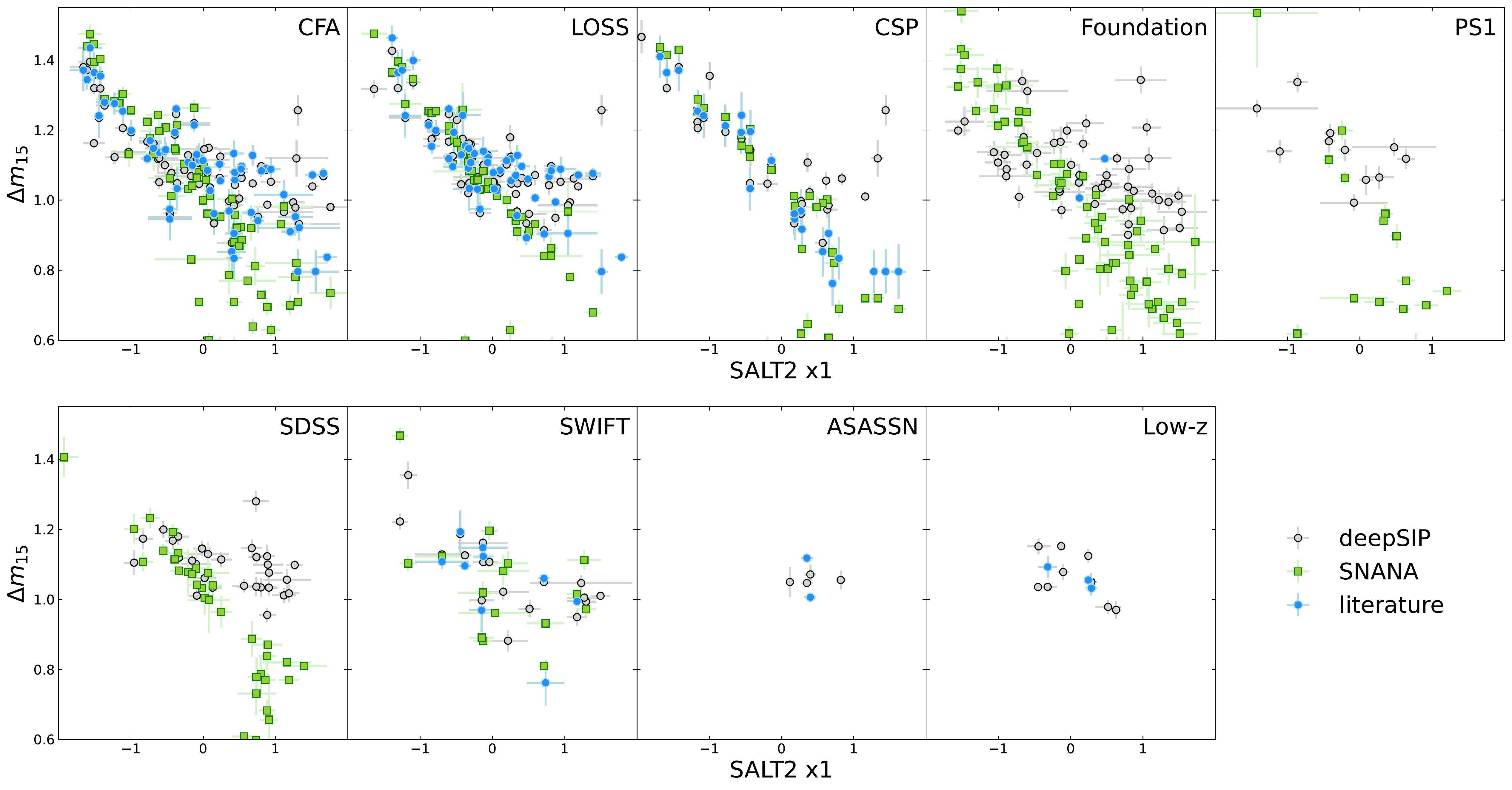}
        \caption{Comparison of \dmfifteen values estimated by \deepSIP and \texttt{SNANA} (\texttt{SNooPy}). The values are compared against \texttt{SALT2} $x_1$ for visual aid. }
        \label{fig:dm15_vs_x1_SNANA}
    \end{figure}

\begin{figure}[t]
        \centering
        \includegraphics[width=\linewidth]{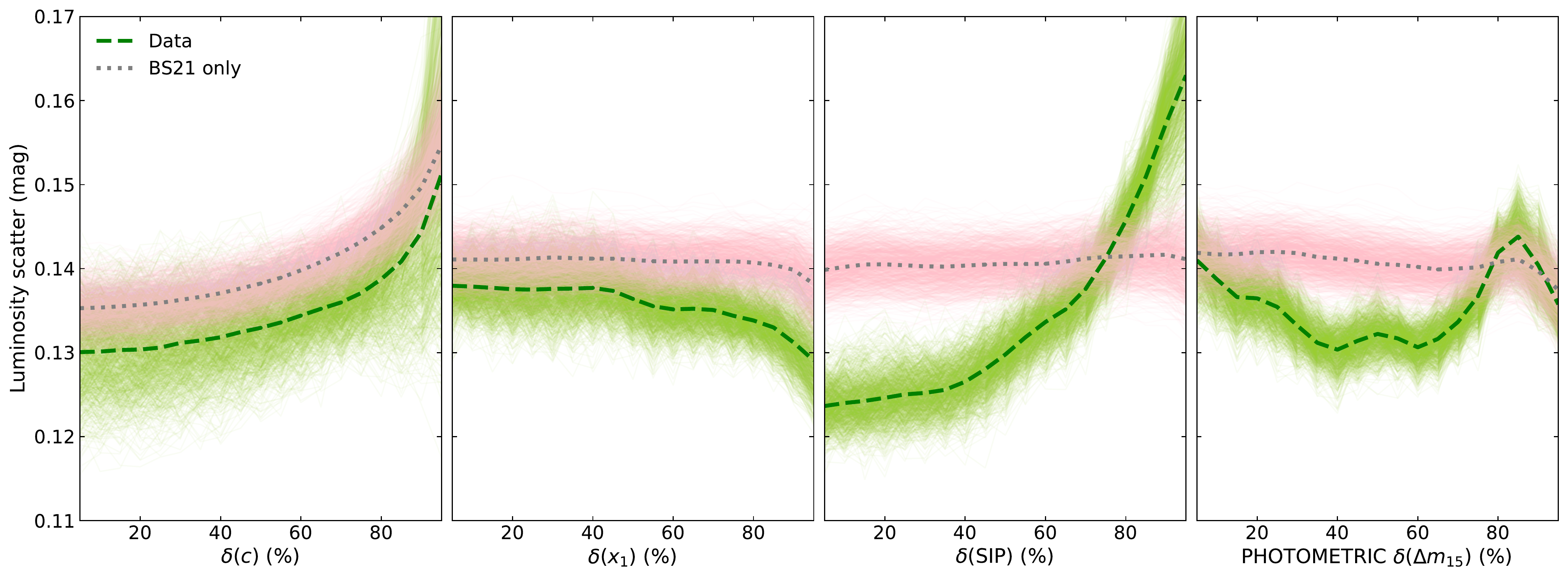}
        \caption{Scatter profile analysis of SN~Ia luminosity using photometric \dmfifteen. The photometrically derived \dmfifteen (the right-most panel) does not exhibit the same trend in scatter profile measured by SIP (second from the right).}
        \label{fig:SP_SNANA}
    \end{figure}

    In \cite{Stahl+2020_deepSIP} and subsequent papers, SIP is presented as \dmfifteen, which is trained to reproduce a photometric parameter estimated by \texttt{SNooPy} \citep{Burns_2011_SNooPy}. Despite this, \dmfifteen measurements produced by \deepSIP rely \emph{solely} on spectra and are therefore purely spectroscopic measurements.
    To provide a comparison of photometrically-derived \dmfifteen values and SIPs, we calculate the \dmfifteen values in a traditional, photometry-only method. 
    In particular, we use the photometric dataset from \texttt{Pantheon+} and the \texttt{SNooPy} model implemented in the \texttt{SNANA} \citep{Kessler_2009_SNANA} package\footnote{A numerical conversion from the ``Stretch" parameter to \dmfifteen \citep{Burns_2014} is needed as SNANA only provides the estimation of Stretch parameter. This conversion does not affect the scatter profile analysis shown in Fig.~\ref{fig:SP_SNANA} since the kernel is translation and scale invariant.}. 
    In Fig.~\ref{fig:dm15_vs_x1_SNANA}, we compare photometrically estimated \dmfifteen values and their corresponding SIP values. When the photometric \dmfifteen values are previously estimated by other works, they are presented as ``literature'' \dmfifteen.
    When this photometric \dmfifteen is used for the scatter profile analysis (similarly to Sec.~\ref{sec:SIP_similarity_trend}), the photometric \dmfifteen value does not show a similar relation as with SIP, as shown in Fig.~\ref{fig:SP_SNANA}. This agrees with the hypothesis that our increased precision is due to the incorporation of information unique to the spectra and not found in the photometry, rather than a simple change in parameterization.
    
    \subsection{Orthogonality of parameters} \label{sec:orthogonality}
    While SIP measured by \deepSIP is shown to provide additional, spectroscopic information (see Sec.~\ref{sec:photometric_dm15}) and SIP and $x_1$ are observationally independent, the value of SIP, which is intended to replicate the light-curve shape \dmfifteen, is similar to the SALT2 $x_1$ parameter and they are both informative about similar underlying properties of each SN. This means that there is a partial correlation between SIP and $x_1$: when a pair of SNe have a similar SIP value, their measured $x_1$ values tend to be similar as well (e.g., Fig.~\ref{fig:dm15_vs_x1_SNANA}). 
    This causes an unwanted upward trend in our scatter profile in $x_1$ when Model 2 is used (see Fig.~\ref{fig:scatter_profile_fitted}). We account for this by including $x_1$ (and $c$) in the $\chi^2$ evaluation, which prevents the model scatter profile from deviating too much from the data in $x_1$ space. This, however, limits the possible size of $\sigma_\text{SIP}$, and thus the model is not fully optimized in the $\delta_\text{SIP}$ space alone. This issue can be resolved when a spectroscopic parameter is orthogonal to (i.e., not correlated with) \texttt{SALT2} parameters. In future work we aim to (i) perform orthogonalization within the current framework, or  (ii) retrain \texttt{deepSIP} and redefine SIP to be an orthogonal parameter, maximizing the information retrieved from all available data.

    \subsection{Number of spectroscopic parameters}
    \citetalias{Boone_2021_TwinsEmbeddinng_I} have suggested that a single spectroscopic parameter may not be enough to capture all spectroscopic variation. \citetalias{Boone_2021_TwinsEmbeddinng_I} found that a three-parameter model most effectively captures the information within the SNfactory sample. Currently our method with \texttt{deepSIP} only provides a single time-independent spectroscopic parameter, and retraining \texttt{deepSIP} to infer additional parameters may enable further reduction of $H_0$ uncertainty, in addition to the benefit mentioned in Sec.~\ref{sec:orthogonality}.
    
\section{Conclusion} \label{sec:conclusion}
    We have determined a SIP --- spectroscopically inferred parameter --- for over 70\% of SH0ES SNe~Ia (100\% Cepheid calibrated and 73\% Hubble flow SNe) using the purpose-built neural network \texttt{deepSIP}. As a spectroscopic property, SIP can be used to improve the standardization of cosmological SNe~Ia beyond the extent possible with existing photometry-based methods.
    We have demonstrated the method of scatter profile evaluation --- i.e., pairwise comparison of observable parameters and the standardized luminosity of SNe~Ia. The scatter profile in $(c, x_1)$ space shows that the current SN~Ia uncertainties are overestimated, and the scatter profile in SIP space shows that spectroscopically similar SNe~Ia tend to have similar luminosities after SALT2 standardization. This result is consistent with existing studies, and the currently employed BS21/P21 model does not capture the SIP-dependent scatter profile.

    Based on the observed scatter profiles, we have built two covariance models --- Model 1 to correct the overestimated uncertainty and Model 2 to account for the SIP-dependent scatter profile. Our new models are provided as a partially modified form of the BS21/P21 model by linearly combining new model matrices to the existing SH0ES SN covariance matrix. The linear coefficients of this model represent the size of the gray scatter to be removed (Model 1) and the size of the SIP-dependent covariance (Model 2) to be added, and their values, as well as an additional scale parameter for Model 2, are determined by minimizing the deviation of scatter profiles between data and the model.

    Using the optimized models, we find that correcting for the overestimated uncertainty (Model 1) reduces the $H_0$ uncertainty by $\sim 5\%$ and yields $H_{0,\, \text{baseline-only}} = 73.14 \pm 0.91$\,km\,s$^{-1}$\,Mpc$^{-1}$. Accounting for the SIP-dependent scatter (Model 2) further reduces the $H_0$ uncertainty by $\sim 7\%$. Taking the systematic error due to choice of hyperparameters and additional systematic error ($\sim0.3$\,km\,s$^{-1}$\,Mpc$^{-1}$) estimated by \citetalias{Riess_2022_SH0ES} into consideration, we report the updated Hubble constant to be $H_0 = 73.29 \pm 0.90$\,km\,s$^{-1}$\,Mpc$^{-1}$. This value remains in high tension against the \citetalias{Planck_2018} value of $H_0 = 67.4 \pm 0.5$\,km\,s$^{-1}$\,Mpc$^{-1}$, now at a $\sim 5.7 \sigma$ level of disagreement.

\acknowledgments
    We thank Yen-Chen Pan for providing the spectra from the Foundation survey. We are grateful to Richard Kessler for providing computing resources and Justin Pierel for suggestions on the SNANA configuration.
    Y.S.M. thanks Wenlong Yuan and Mi Dai for discussions of covariance matrices and Gaussian process, and Ken Takaki for the inspiration in the measurement methods. 
    A.V.F.'s and B.E.S.'s work has been financed by the Christopher R. Redlich Fund and many individual donors.

    Funding for SDSS and SDSS-II has been provided by the Alfred P. Sloan Foundation, the Participating Institutions, the National Science Foundation (NSF), the U.S. Department of Energy, the National Aeronautics and Space Administration (NASA), the Japanese Monbukagakusho, the Max Planck Society, and the Higher Education Funding Council for England. The SDSS Web Site is \url{http://www.sdss.org/}.
    The SDSS is managed by the Astrophysical Research Consortium for the Participating Institutions. The Participating Institutions are the American Museum of Natural History, Astrophysical Institute Potsdam, University of Basel, University of Cambridge, Case Western Reserve University, University of Chicago, Drexel University, Fermilab, the Institute for Advanced Study, the Japan Participation Group, Johns Hopkins University, the Joint Institute for Nuclear Astrophysics, the Kavli Institute for Particle Astrophysics and Cosmology, the Korean Scientist Group, the Chinese Academy of Sciences (LAMOST), Los Alamos National Laboratory, the Max-Planck-Institute for Astronomy (MPIA), the Max-Planck-Institute for Astrophysics (MPA), New Mexico State University, Ohio State University, University of Pittsburgh, University of Portsmouth, Princeton University, the United States Naval Observatory, and the University of Washington.

    The Pan-STARRS1 Surveys (PS1) and the PS1 public science archive have been made possible through contributions by the Institute for Astronomy, the University of Hawaii, the Pan-STARRS Project Office, the Max-Planck Society and its participating institutes, the Max Planck Institute for Astronomy, Heidelberg and the Max Planck Institute for Extraterrestrial Physics, Garching, The Johns Hopkins University, Durham University, the University of Edinburgh, the Queen's University Belfast, the Harvard-Smithsonian Center for Astrophysics, the Las Cumbres Observatory Global Telescope Network Incorporated, the National Central University of Taiwan, the Space Telescope Science Institute, NASA under grant NNX08AR22G issued through the Planetary Science Division of the NASA Science Mission Directorate, NSF  grant AST-1238877, the University of Maryland, Eotvos Lorand University (ELTE), the Los Alamos National Laboratory, and the Gordon and Betty Moore Foundation.
    
\appendix
\section{Understanding and avoiding bias in SIP due to spectral truncation} \label{sec:dm15-cut-bias}

    As described in Sec.~\ref{sec:deepSIP-preprocessing}, the normalized spectra of SNe~Ia after initial preprocessing can contain significant noise near the 7000\,\AA\ region. Such excessive noise could lead to \deepSIP falsely classifying SNe~Ia as ``out-of-domain'' (Sec.~\ref{sec:deepSIP}), even when the physical properties of the given spectra (i.e., Phase and \dmfifteen) are within the defined domain. 
    Removing the noisy part of spectra is an effective way of correcting the false classification of the domain output from \deepSIP, and we systematically perform iterative cuts, both on the blue and red sides, to check if any of such cuts yield ``in-domain'' classification to each spectrum.
    Great caution was taken when choosing the combination of blue and red cuts, as doing so could lead to a false deviation (bias) of the estimated physical parameters (Phase, \dmfifteen, and their associated uncertainties). To choose the range of cuts that limits the possible bias within our target ($\sim 1\sigma$ deviation from the truth value), we performed an extensive validation test.

    First, we choose a set of high-quality spectra (hereafter the validation set) whose physical properties are within the domain. We confirm that all validation-set spectra can be processed with \deepSIP without cuts and achieve an ``in-domain'' classification. We record the \deepSIP output set (ii--v; see Sec.~\ref{sec:deepSIP}) of the unaltered spectra as truth values of the \dmfifteen and Phase. 
    We then apply wavelength cuts to the red and blue ends of the spectra. Cuts on the blue end are applied from 3400\,\AA\ to 5000\,\AA\ (the Fe~II line), while cuts on the red end are applied from 6000\,\AA\ (the Si~II line) to 7500\,\AA. The region between 5000\,\AA\ and 6000\,\AA\ is excluded because it contains crucial SN~Ia features. 
    Spectra with varying cuts are processed by \deepSIP to obtain their new \dmfifteen and Phase values. 
     \begin{figure}
        \centering
        \includegraphics[width=\linewidth]{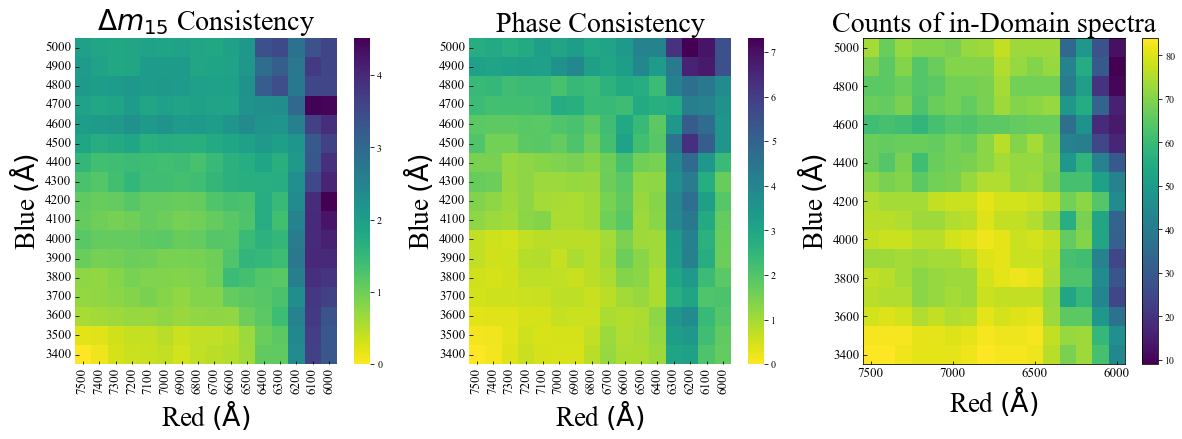}
        \caption{Validation results for noise-cut wavelengths and yielded physical parameters, averaged over all ``validation set'' spectra. Lighter, yellow colors indicate smaller mean bias introduced by the corresponding set of blue and red cuts, and darker, bluer colors indicate significant deviations due to the chosen cuts, which should be avoided. {\it Left:} deviation in \dmfifteen values. {\it Middle:} deviation in Phase values. {\it Right:} counts of in-domain spectra.}
        \label{fig:bias-analysis}
    \end{figure}
        
    The mean deviation of resulting \dmfifteen and Phase values from the truth, as well as the counts of ``in-domain'' classification, are shown in Fig.~\ref{fig:bias-analysis}. The deviation is calculated in the unit of uncertainty (i.e., the standard ``sigma'' deviation). Our results suggest that the red-side cut has a significant effect on the estimated physical parameters near 6100\,\AA\ (Si~II lines), and the blue-side cut has a rather mild slope of deviation. A similar trend is also evident in the Phase space.
    We determine that any cuts between 4400 and 6300\,\AA\  yield biases, with the most frequent and significant occurring near the Fe~II and Si~II lines. To limit the mean deviation within $1\sigma$ of the truth value, we define our range of possible combinations for noise cuts based on the region of least bias: between 3400\,\AA\ and 4400\,\AA\ on the blue end of the spectrum and between 6300\,\AA\ and 7500\,\AA\ on the red end. We also observe fewer occurrences of falsely classified ``out-of-domain'' outputs in the selected range of wavelength cuts.

\bibliographystyle{JHEP}
\bibliography{main.bib}

\end{document}

%% file: spec_sources.tex
\begin{table}
    \small
    \setlength{\tabcolsep}{3pt}
    \centering
    \caption{Sources of the SN~Ia spectra used in this work.$^a$}
    \begin{tabular}{l|ccccccc|cc}
        \hline
        Source of Spectra & CFA$^b$ & LOSS$^c$ & CSP & SDSS & Foundation & PS1 & Other$^d$ & $N^\text{unique}_\text{SN}$ & $N_\text{spec}$ \\
        \hline
        \citetalias{Blondin_2012_CFAspec}        &   60 &    38 &   14 &     0 &           0 &    0  &      9 &          75 &     676\\
        \citetalias{Matheson_2008_CFAspec_indiv} &    6 &     3 &    0 &     0 &           0 &    0  &      1 &           7 &      92\\
        \citetalias{Stahl_2020_BSNIP}            &    6 &     9 &    6 &     0 &           7 &    0  &     11 &          24 &     121\\
        \citetalias{Silverman_2012_BSNIP1}       &   53 &    40 &   19 &     1 &           0 &    0  &     10 &          75 &     284\\
        \citetalias{Folatelli_2013_CSPspec}      &   15 &    12 &   25 &     0 &           0 &    0  &      3 &          28 &     202\\
        \citetalias{Sako_2018_SDSS-SN-DR}        &    0 &     0 &    0 &    37 &           0 &    0  &      0 &          37 &      72\\
        \citetalias{Dettman_2021_FoundationSpec} &    0 &     1 &    1 &     0 &          23 &    0  &      2 &          24 &      25\\
        \citetalias{Kenworthy_2021_SALT3}        &    0 &     0 &    0 &     0 &          47 &    0  &      0 &          47 &      47\\
        \citetalias{Pan_2022_PS1Spec}            &    0 &     0 &    0 &     0 &           0 &   17  &      0 &          17 &      17\\
        \citetalias{Riess_2022_SH0ES}$^e$    &    0 &     3 &    2 &     0 &           1 &   0  &      5 &           8 &      21\\
        OSC (\citetalias{Guillochon_2017_OSC})    &    1 &     2 &    1 &     0 &           7 &    0  &      6 &          15 &      76\\
        \hline
        Spectra availability (per $N^\text{phot}_\text{SN}$) & 75/79& 58/60 & 37/39& 37/37 &      63/76  & 17/20 & 35/47  &  \textit{249/285} & \\
        \hline
    \end{tabular}
    
    \footnotesize{
    $^a$The left column shows the publications providing the spectroscopic data, most of which are the official data releases from specific surveys. The middle columns list the number of photometric SNe~Ia in the \texttt{SH0ES} baseline sample covered by each publication. The counts are separated into major surveys that comprise the \texttt{SH0ES} data. It should be noted that these values can contain duplicates when multiple independent photometric data are listed for a single SN. The right column represents the number of unique (i.e., not counting photometric duplicates) SNe in the \texttt{SH0ES} baseline that each source covers, as well as the total number of spectra obtained. The bottom row reports the number of SNe listed in each survey whose spectra are obtained. The number printed in the \textit{italic} font at the bottom right represents the total number of unique SNe~Ia whose spectra are obtained.\\
    $^b$Combination of all CFA surveys. \\
    $^c$Photometric objects observed by the Lick/KAIT telescope only. SNe monitored by CFA are marked as \texttt{CFA}. \\
    $^d$Combination of \texttt{Swift}, \texttt{ASASSN}, and \texttt{LOWZ} defined in the \texttt{SNANA} framework.\\
    $^e$Carried over from \citetalias{Riess_2022_SH0ES} when a calibrator SN's spectra are not included in the major data releases (rows above).}
 
    \label{tab:spec_sources}
\end{table}